\documentclass[twocolumn]{aa}
\usepackage{color}
\usepackage{epsfig}
\usepackage{soul}
\usepackage{float}
\usepackage{enumitem}
\usepackage{appendix}
\usepackage{wrapfig}
\usepackage{graphicx}
\usepackage{multicol}
\usepackage{hyperref}
\usepackage{chemfig}
\usepackage[version=3]{mhchem}
\hypersetup{
    colorlinks=true,
    linkcolor=blue,
    filecolor=magenta,
    citecolor=blue,
    urlcolor=blue,
    pdftitle={Overleaf Example},
    pdfpagemode=FullScreen,
    }
\raggedcolumns
\usepackage{natbib}
\def\gtsim{{_>\atop{^\sim}}}
\def\ltsim{{_<\atop{^\sim}}}

\def\aap{A\&A}
\def\apjs{ApJS}
\def\hcopp{HC$^{17}$O$^+$}

\begin{document}

\title{Tracing the contraction of the pre-stellar core L1544 with HC$^{17}$O$^+$ $J$ = 1 - 0 emission.\thanks{Based on observations carried out with the IRAM 30m Telescope. IRAM is supported by INSU/CNRS (France), MPG (Germany) and IGN (Spain)}}

\author{J.~Ferrer Asensio\inst{1}, S.~Spezzano\inst{1}, P.~Caselli\inst{1}, F. O.~Alves\inst{1}, O.~Sipil{\"a}\inst{1}, E.~Redaelli\inst{1}, L.~Bizzocchi\inst{1,2,3}, F. Lique\inst{4}, A. Mullins\inst{1}}
\institute{\tiny{\inst{1} Centre for Astrochemical Studies, Max-Planck-Instiut f\"ur extraterrestrische Physik, Giessenbachstr. 1, 85748 Garching, Germany\\ \inst{2} Scuola Normale Superiore, Piazza dei Cavalieri 7, 56126 Pisa, Italy \\ \inst{3} Dipartimento di Chimica "G. Ciamician", via F. Selmi 2, 40126 Bologna, Italy \\ \inst{4} IPR, Université de Rennes, Bât 11b, Campus de Beaulieu, 263 avenue du Général Leclerc
35042 Rennes Cedex, France\\}}
\date{Received  ; accepted }

\abstract
{Spectral line profiles of several molecules observed towards the pre-stellar core L1544 appear double-peaked. For abundant molecular species this line morphology has been linked to self-absorption. However, the physical process behind the double-peaked morphology for less abundant species is still under debate.  }
{In order to understand the cause behind the double-peaked spectra of optically thin transitions and their link to the physical structure of pre-stellar cores, we present high-sensitivity and high-spectral resolution \hcopp $J =$1-0 observations towards the dust peak in L1544.}
{We observed the \hcopp (1-0) spectrum with the Institut de Radioastronomie Millim\'etrique (IRAM) 30m telescope. By using new state-of-the-art collisional rate coefficients, a physical model for the core and the fractional abundance profile of \hcopp, the hyperfine structure of this molecular ion is modelled for the first time with the radiative transfer code \textsc{loc} applied to the predicted chemical structure of a contracting pre-stellar core. We applied the same analysis to the chemically related C$^{17}$O molecule.}
{The observed \hcopp(1-0) and C$^{17}$O(1-0) lines have been successfully reproduced with a non-local thermal equilibrium (LTE) radiative transfer model applied to chemical model predictions for a contracting pre-stellar core. An upscaled velocity profile (by 30\%) is needed to reproduce the \hcopp(1-0) observations.  }
{The double peaks observed in the \hcopp(1-0) hyperfine components are due to the contraction motions at densities close to the critical density of the transition ($\sim$10$^{5}$ cm$^{-3}$) and to the fact that the HCO$^{+}$ fractional abundance decreases toward the centre.}

\authorrunning{J. Ferrer Asensio et al.}
\titlerunning{HC$^{17}$O$^+$ hyperfine emission in L1544: Observations and radiative transfer modelling.}

\maketitle
\section{Introduction} \label{introduction}

Pre-stellar cores are gravitationally bound cores seen as sub-structures within molecular clouds. They present high central densities ($n_{H_2}> 10^{5}$ cm$^{-3}$) and low temperatures at the centre ($<10$ K) \citep{keto:08}. These sources are on the verge of contraction but have yet not formed a protostar. The combined study of the physical structure and kinematics of pre-stellar cores, which represent the earliest stages of star formation \citep{andre:14}, is crucial to achieve a comprehensive view of the initial conditions for core collapse. Molecular spectra have been widely used as diagnostics of dense cloud cores. The intensities, widths, rest frequencies and profiles of spectral lines allow to characterise the physical, chemical and kinematic structure of the source they originate from. Transitions with hyperfine structure are especially useful for deriving physical properties such as optical depth and excitation temperature ($T_\mathrm{ex}$) of the line emitting area. The column density can then be derived without the need of assumptions on $T_\mathrm{ex}$ or of observations of other transitions, making hyperfine transitions useful probes of the physics and chemistry of molecular clouds (e.g. \citealt{caselli:95, sohn:07, lique:15}).\

L1544 is a widely studied pre-stellar core in the Taurus Molecular Cloud at a distance of 170 \,pc \citep{galli:19}.  Its chemical composition and structure have been constrained in multiple studies \citep{tafalla:98, caselli:02, crapsi:05, keto:101, vastel:14, spezzano:17, caselli:19, chacon:19, jin:20}. L1544 shows signs of contraction motions, which resemble the quasi-equilibrium contraction of a Bonnor Ebert sphere \citep{keto:15}. The core is also centrally concentrated, with central H$_{2}$ column densities close to 10$^{23}$ cm$^{-2}$ \citep{ward:99, crapsi:05}, low central temperatures ($\sim$ 7 K, \citealt{crapsi:07}), large amount of freeze-out \citep{caselli:22}, large deuterium fractions \citep{crapsi:05, redaelli:19} and a rich chemical composition \citep{vastel:14, jimenez:16, spezzano:17}. Due to its characteristics, L1544 is an ideal source to study the dynamics of a dense core on the verge of star formation.\

Previous observations towards the dust peak of L1544 recorded spectral line transitions that present a double-peaked structure \citep{tafalla:98, caselli:99, williams:99, dore:01, caselli:02}. This line morphology can be explained for some cases, as for example the N$_{2}$H$^{+}$(1-0) line \citep{williams:99}, by self-absorption from a less dense, contracting envelope \citep{keto:10}. Self-absorption is expected to arise for optically thick transitions of abundant molecules \citep{tafalla:98}. However, lines of rare isotopologues, such as D$^{13}$CO$^{+}$(1-0) and HC$^{17}$O$^{+}$(1-0), are expected to be optically thin and their double-peaked profile is not expected to arise from self-absorption. Possible reasons for these features are the presence of dense material at different velocities along the line of sight \citep{tafalla:98} or the depletion of the targeted molecules towards the inner part of the core \citep{caselli:02}. \

In order to unveil the nature of the process responsible for double-peaked line profiles of optically thin species, we obtained new high-sensitivity observations of the HC$^{17}$O$^{+}$(1-0) line toward L1544. The hyperfine structure of the line occurs because the nuclear spin of $^{17}$O ($I$ = 5/2) and the molecular rotation are coupled, resulting in the splitting of the $J$ = 1 rotational level. The reason for choosing this particular isotope of HCO$^{+}$ is that it is expected to have optically thin lines that should not be affected by self-absorption and will allow to see across the totality of the core. \hcopp \ was first observed in the interstellar medium by \cite{guelin:82} towards Sagittarius B2 with the Bell Laboratories (BTL) 7m telescope at Crawford Hill Laboratory, New Jersey. They detected the $J$ = 1 - 0 rotational transition centred at 87.1 GHz with an angular resolution of 2\arcmin{} and spectral resolution of 1 MHz (3.5 km s$^{-1}$). \hcopp(1-0) has been detected more recently with the Institut de Radioastronomie Millim\'etrique (IRAM) 30m telescope towards the dust peak in L1544 (we refer to Figure 2 in \cite{dore:01} and Figure 1 in \cite{caselli:02}).\

In this paper we present a new high-sensitivity and high-spectral resolution HC$^{17}$O$^{+}$(1-0) spectrum towards the dust peak of L1544 observed with the IRAM 30m telescope. With the aid of predictions from a chemical model, applied to the physical structure of L1544, and the non- local thermodynamic equilibrium (LTE) radiative transfer code \textsc{loc}, we reproduce the observed spectrum and discuss the results. This article is structured as follows: in Section \ref{observations} we describe the observational details of the data obtained with the IRAM 30m telescope; in Section \ref{radiative} we present the determination of the collisional rate coefficients required for the modelling and we describe the non-LTE radiative transfer code, the pre-stellar core model and the fractional abundance profiles used. The results are presented in Section \ref{results}. In Section \ref{discussion} we discuss the results obtained in the context of past works, and our conclusions can be found in Section \ref{conclusions}.\\

\section{Observations} \label{observations}

\begin{figure}[H]
\centering
\includegraphics[width=10cm]{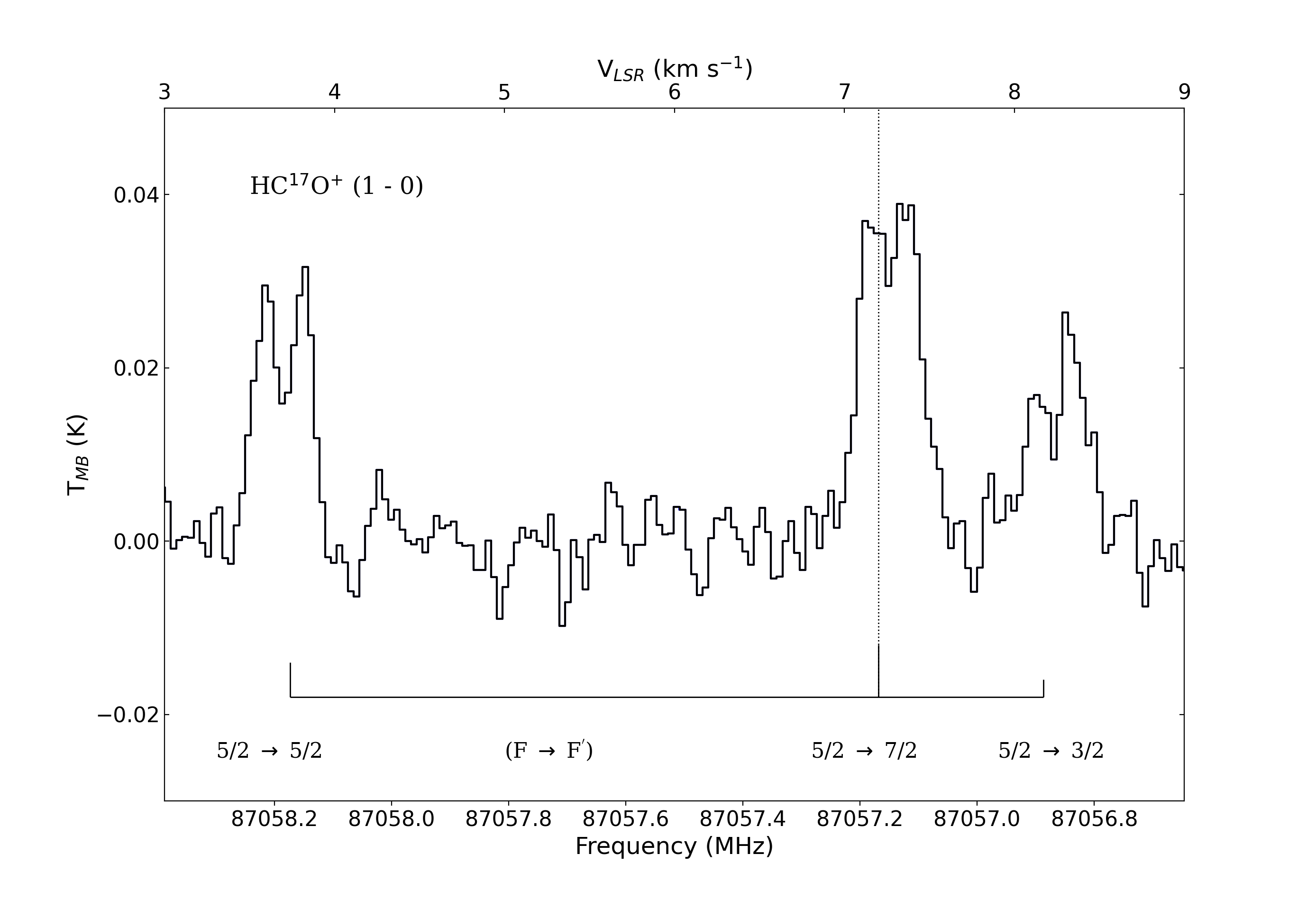}
\caption{Spectrum of \hcopp(1-0) at the dust peak of L1544. The hyperfine structure is shown by vertical solid lines with heights proportional to their relative intensities (see Table \ref{tab}). The vertical dotted line represents the LSR velocity of L1544 (7.2 km s$^{-1}$).}
\label{hc17o+}
\end{figure}

We observed the ground state $(J = 1 - 0)$ rotational transition of \hcopp \ in a single pointing towards the dust peak of L1544 ($\alpha _{2000}$ = 05$^h$04$^m$17$^s$.21,  $\delta _{2000}$ = +25$^\circ$10$'$42$''$.8). These observations were carried out from Oct 10th to 15th 2018 using the IRAM 30m telescope at Pico Veleta, with a total on-source integration time of 22.6~hr. The telescope pointing was checked frequently against the nearby bright quasar B0316+413 and found to be accurate within 4". We used the E090 band of the EMIR receiver. The Half Power Beam Width (HPBW) of these observations is 28". The tuning frequency was 87.060~GHz and the observations were performed in frequency switching mode. We used the VESPA backend to achieve a spectral resolution of 10~kHz, equivalent to $\Delta v$ = 0.034~km s$^{-1}$ at this frequency. Both horizontal and vertical polarisations were observed simultaneously. The final spectrum is shown in Figure \ref{hc17o+}.\ 

The observational data was processed and then averaged with \textsc{class} (Continuum and Line Analysis Single-dish Software), an application from the \textsc{gildas}\footnote{\url{https://www.iram.fr/IRAMFR/GILDAS/}} software \citep{pety:05}. A F$_{eff}$/B$_{eff}$ ratio of 1.17 was used for the $T_{A}^{*}$ to $T_\mathrm{MB}$ conversion.

Figure \ref{hc17o+} shows the observed \hcopp(1-0) rotational transition split into three hyperfine components which present a clear double-peaked structure. Each hyperfine level is 
labelled by a quantum number $F$ ($\mathbf{F} = \mathbf{I} + \mathbf{J}$) varying between $\left|I-J\right|$ and $I + J$\@. The quantum number $F$ corresponds to the lower level while the $F^{\prime}$ corresponds to the upper level of the transition.  Table \ref{tab} reports the hyperfine transition frequencies measured in the laboratory, relative intensities and upper energies. A signal-to-noise ratio (S/N) of 7 was obtained for the least bright component with a rms of 2.8 mK. The intensity difference between the two peaks in the $F - F^{'}$ = 5/2 $\rightarrow$ 3/2 transition centred at a frequency of 87.057 GHz is comparable to the noise level.
\begin{table}[H]
\begin{center}
\caption{Hyperfine transition frequencies measured in the laboratory \citep{dore:01}, upper energies of \hcopp(1-0) and relative intensities.}
\begin{tabular}{ cccc } 
\hline\hline
$F$ $\rightarrow$ $F^{'}$ & Frequency &  $E_\mathrm{up}$ & Relative\\
 & (MHz) & (K) & intensity \\
\hline
5/2$\rightarrow$3/2 & 87056.966(20) & 4.18 & 2\\
5/2$\rightarrow$7/2 & 87057.258(20) & 4.18 & 4\\
5/2$\rightarrow$5/2 & 87058.294(20) & 4.18 & 3\\
 \hline
\label{tab}
\end{tabular}
\end{center}
\end{table}

Moreover, we use C$^{17}$O(1-0) observations towards the dust peak of L1544 from \cite{chacon:19}, for which an On-The-Fly (OTF) map was taken with the IRAM 30m telescope with a spectral resolution of 20 kHz ($\Delta v$ = 0.05 km s$^{-1}$). For C$^{17}$O, a single pointing spectrum was extracted towards the dust peak within a beam of 22" which corresponds to the IRAM 30m beam at the C$^{17}$O(1-0) transition frequency, 112.360 GHz. A F$_{eff}$/B$_{eff}$ ratio of 1.20 was used for the T$_{A}^{*}$ to $T_\mathrm{MB}$ conversion. The resulting spectrum has a rms of 75 mK (See Section \ref{c17o}).\ 

\section{Radiative transfer} \label{radiative}

In order to fully understand the observed spectra, we need to take into account the transfer of radiation across the source. Physical conditions can vary along the radiation's path which will affect the observed line intensity and profile. To derive information carried by spectral line profiles, we need to carry out a full radiative transfer modelling of the emission. Radiative transfer accounting for the physical structure of the observed object is crucial for the correct interpretation of spectra, as it has been shown in past works, including those on pre-stellar cores \citep{caselli:02, sohn:07, keto:15, redaelli:19}.\ 

The present work focuses on L1544, a pre-stellar core with well-studied density, temperature and velocity profiles \citep{ward:99,crapsi:05,crapsi:07,keto:15}. In such environments, LTE conditions do not generally apply across the source. In LTE, the energy level populations of molecules are described by the Boltzmann distribution. LTE applies when the critical density of the transition ($n_{crit}$= $A_{ul}$/$k_{ul}$, where $A_{ul}$ is the Einstein A coefficient and $k_{ul}$ is the collisional coefficient where "u" and "l" are the upper and lower levels respectively) is equal or lower than the volume density of the emitting region. In pre-stellar cores, volume densities can range from $\gtsim$10$^{6}$ cm$^{-3}$ at the centre to $\ltsim$10$^{2}$ cm$^{-3}$ at the edge. For example, the CO 1-0 transition has a n$_{crit}$ of about 10$^{3}$ cm$^{-3}$, making LTE not applicable in the outer parts of the core. For transitions with high n$_{crit}$, LTE applies only in a small region near the centre of the core. When departing from LTE, collisional rate coefficients are needed to study the transfer of radiation. Furthermore, in non-LTE conditions, the hyperfine component intensity ratio may diverge from statistical weights (e.g. \citealt{caselli:95, bizzocchi:13, Faure:12, mullins:16}). In order to accurately predict line intensities resulting from different physical conditions, a full non-LTE radiative transfer treatment with accurate collisional rate coefficients is required.\ 

In Section \ref{hyper} we describe the methodology for the hyperfine collisional rate coefficients calculations, in Section \ref{loc} we describe the physical model adopted for L1544 and the \textsc{loc} radiative transfer code and in Section \ref{chemical} the chemical code used to compute the molecular fractional abundance profile.

\subsection{Hyperfine collisional rate-coefficient calculations} \label{hyper}
We have determined HC$^{17}$O$^+$--H$_2$ rate coefficients from the Close Coupling (CC) HCO$^+$--H$_2$ rate coefficients ($k^{CC}_{J \to J'} (T)$) of \cite{Yazidi:14} using the Infinite Order Sudden (IOS) approximation described in \cite{Faure:12}. We use HCO$^+$--H$_2$ collisional rate coefficients for \hcopp as we do not expect a significant difference, similarly as seen in \cite{daniel:16} where the collisional rate-coefficients for N$_{2}$H$^{+}$ (isoelectronic of HCO$^{+}$) and its $^{15}$N isotopologues are been found to be similar.

In HC$^{17}$O$^+$, the coupling between the nuclear spin ($I=5/2$) of the $^{17}$O atom and the molecular rotation results in a weak splitting \citep{alexander:85} of each rotational level $J$, into hyperfine levels. Each hyperfine level is designated by a quantum number $F$ ($F=I+J$) varying between $|I-J|$ and $I+J$. 

Within the IOS approximation, inelastic rotational
rate coefficients $k^{IOS}_{J \to J'}(T)$ can be calculated from the ''fundamental'' rates (those out of the lowest $J=0$ level) as follows \citep{corey83}:
\begin{equation}
k^{IOS}_{J \to J'}(T)=(2J'+1)\sum_L \left(\begin{array}{ccc}
J' & J & L \\ 
0 & 0 & 0
\end{array}\right)^{2} k^{IOS}_{0 \to L}(T).
\label{iosrot}
\end{equation}

Similarly, IOS rate
coefficients among hyperfine structure levels can be obtained from the
$k^{IOS}_{0 \to L} (T)$ rate coefficients using the following formula
\citep{corey83}:

\begin{eqnarray} \label{REEQ}
\begin{aligned}
& k^{IOS}_{jF \to j'F'} (T)  =
(2j+1)(2j'+1)(2F'+1)  \sum_{L} \left(\begin{array}{ccc}
j' & j & L \\ 
0 & 0 & 0
\end{array}\right)^{2}
\times\\
& \left\{\begin{array}{ccc}
j & j' & L \\
F' & F & I 
\end{array}
\right\}^2 k^{IOS}_{0 \to L} (T). 
\end{aligned}
\end{eqnarray}

\noindent
In the above, $\left( \quad \right)$ and $\left\{ \quad \right\}$ are
respectively the ``3-j'' and ``6-j'' Wigner symbols.\\

The IOS approximation is however expected to be only moderately accurate at low temperature. As a result, we have computed the hyperfine rate coefficients as follows \citep{Faure:12}:

\begin{equation}
\label{scaling}
  k^{INF}_{JF \to J'F'} (T) = \frac{k^{IOS}_{JF \to J'F'}(T)}{k^{IOS}_{J\to
    J'}(T)}k^{CC}_{J\to J'}(T),
\end{equation}
using the CC rate coefficients $k^{CC}(0\to L)$ of \cite{Yazidi:14} for the IOS
``fundamental'' rates ($k^{IOS}_{J\to J'}(T)$) in
Eqs.~\ref{iosrot}-\ref{REEQ}. 

In addition, we note that the fundamental excitation rates
$k^{CC}_{0\to L}$ were replaced by the de-excitation
fundamental rates using the detailed balance relation:
\begin{equation}
k^{CC}_{0\to L}=(2L+1)k^{CC}_{L\to 0}. \, 
\end{equation}
This procedure is found to significantly improve the results
at low temperature due to important threshold effects.\

The calculated rate coefficients can be found in Appendix \ref{colcoe}.

\subsection{The \textsc{loc} radiative transfer code} \label{loc}

Using the radiative transfer code \textsc{loc} (line transfer with OpenCL \cite{juvela:20}), we calculate the strength and shape of the \hcopp(1-0) spectrum taking into account its hyperfine structure. The level populations are calculated with the statistical equilibrium equations, which are solved using an Accelerated Lambda Iteration (ALI) \citep{rybicki:91} and the molecule's radiative and collisional rates. The radiative transfer is calculated using the 1D model in \textsc{loc} which takes into account the volume density, the kinetic temperature, radial velocity and micro-turbulence of a spherically symmetric physical structure (see below for pysical parameter ranges).\

The physical parameters for the modelling are extracted from a pre-stellar core physical model based on \cite{keto:15}. This one-dimensional model of an unstable quasi-equilibrium Bonnor-Ebert sphere has been previously shown to be successful in reproducing the profile of several molecular transitions observed towards the dust peak in L1544 \citep{keto:10, keto:101, caselli:12, bizzocchi:13, caselli:17, redaelli:18, redaelli:19, redaelli:21, caselli:22}.\ 

\begin{figure}[H]
\centering
\includegraphics[width=9cm]{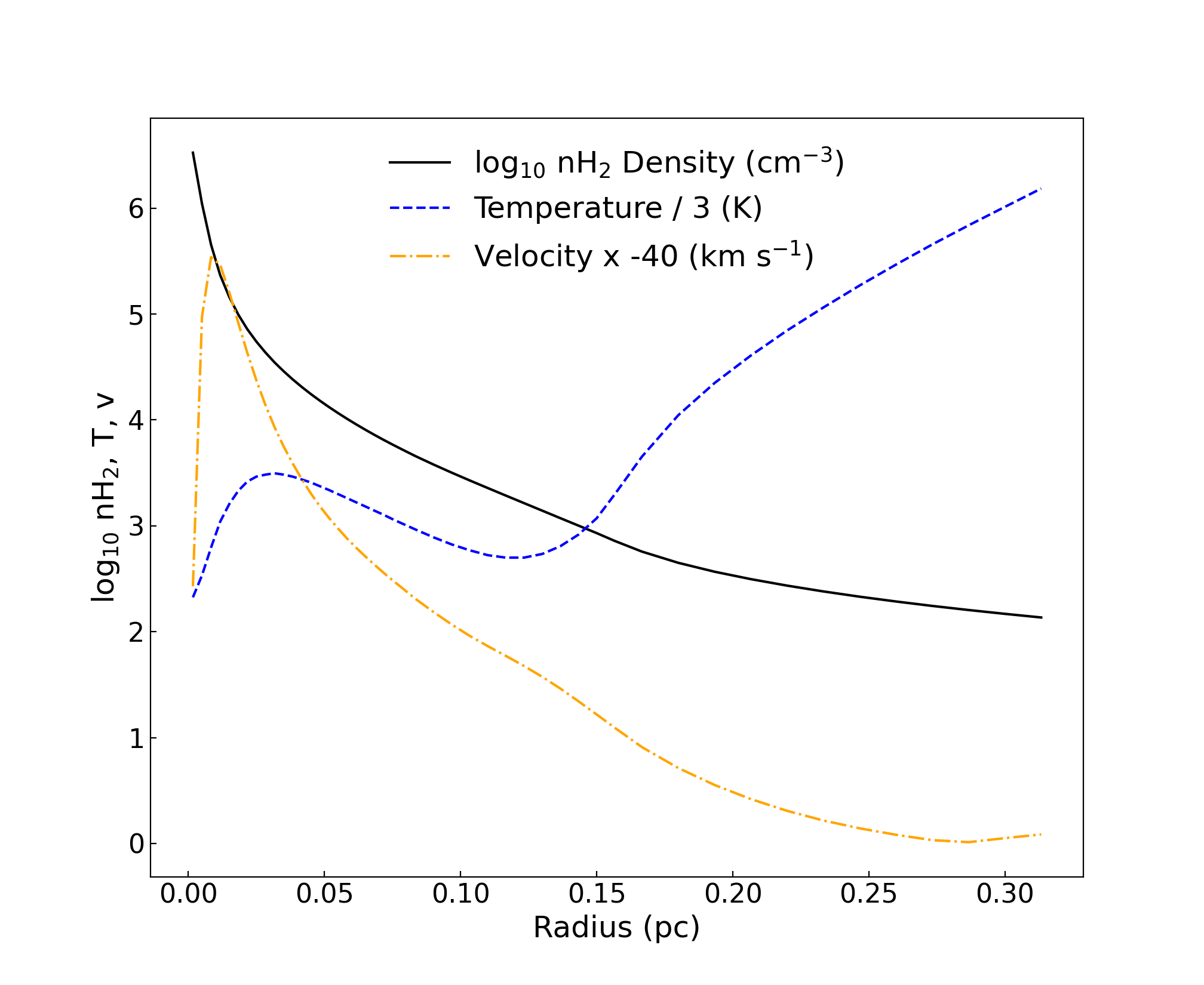}
\caption{The black solid line represents the number density of molecular hydrogen, the blue dashed line the gas temperature, and the orange dashed and dotted line the velocity profile of the \cite{keto:15} physical model. In order to display the various parameters in a single plot, we are showing the base 10 logarithm of the number density of molecular hydrogen, the temperature divided by 3 and the velocity scaled by -40.}
\label{pm}
\end{figure}

The radial profile of physical properties of the model are plotted in Figure \ref{pm}.We show in this plot the molecular hydrogen number density ($n_{H_2}$, black solid line), the gas temperature ($T$, blue dashed line) and the velocity ($v$, orange dashed and dotted line). Notice that, in order to include these three parameters in this one plot we show the logarithm of the molecular hydrogen number density, the gas temperature divided by 3 and the velocity scaled by -40. The n($H_2$) ranges from 8$\times10^{6}$ cm$^{-3}$ at the centre of the core to 1$\times10^{2}$ cm$^{-3}$ at the edge of the core (0.32 pc). $T$ ranges from 6 K at the centre of the core to 18 K at the edge. Lastly, $v$ ranges from -0.14 km s$^{-1}$ at the velocity peak ($\sim$ 0.01 pc) to -0.01 km s$^{-1}$ at the edge of the core.

\subsection{Chemical code} \label{chemical}

The fractional abundance profile of \hcopp \, is simulated using the pseudo-time dependent gas-grain chemical model, described in \cite{sipila:15}. We use here the same approach to simulate the chemical fractional abundances in L1544 as laid out in \cite{sipila:19}. Briefly, we adopt the physical model for L1544 of \cite{keto:15} which is separated into concentric shells; the results of chemical simulations carried out in each shell are combined, yielding time-dependent fractional abundance profiles for the various molecules in the chemical network. The physical structure of the core is thus fixed, while the chemistry evolves in a time-dependent manner. We refer the reader to \cite{sipila:15,sipila:19} for more details on the model, for example the initial fractional abundances. For the present work we extract several HCO$^{+}$ fractional abundance profiles in a time interval between 10$^{4}$ and 10$^{7}$ years. The model does not treat oxygen isotope chemistry; to obtain the \hcopp \, fractional abundance, we scale the HCO$^{+}$ fractional abundance by the isotopic ratio [$^{16}$O/$^{17}$O] = 2044 \citep{penzias:81, wilson:94}. The same procedure is also followed for C$^{17}$O.\ 

\section{Results} \label{results}
\subsection{\texorpdfstring{\hcopp}{HC17O+}}
\label{shc17o+}

\begin{figure*}[th]
\includegraphics[width=\textwidth]{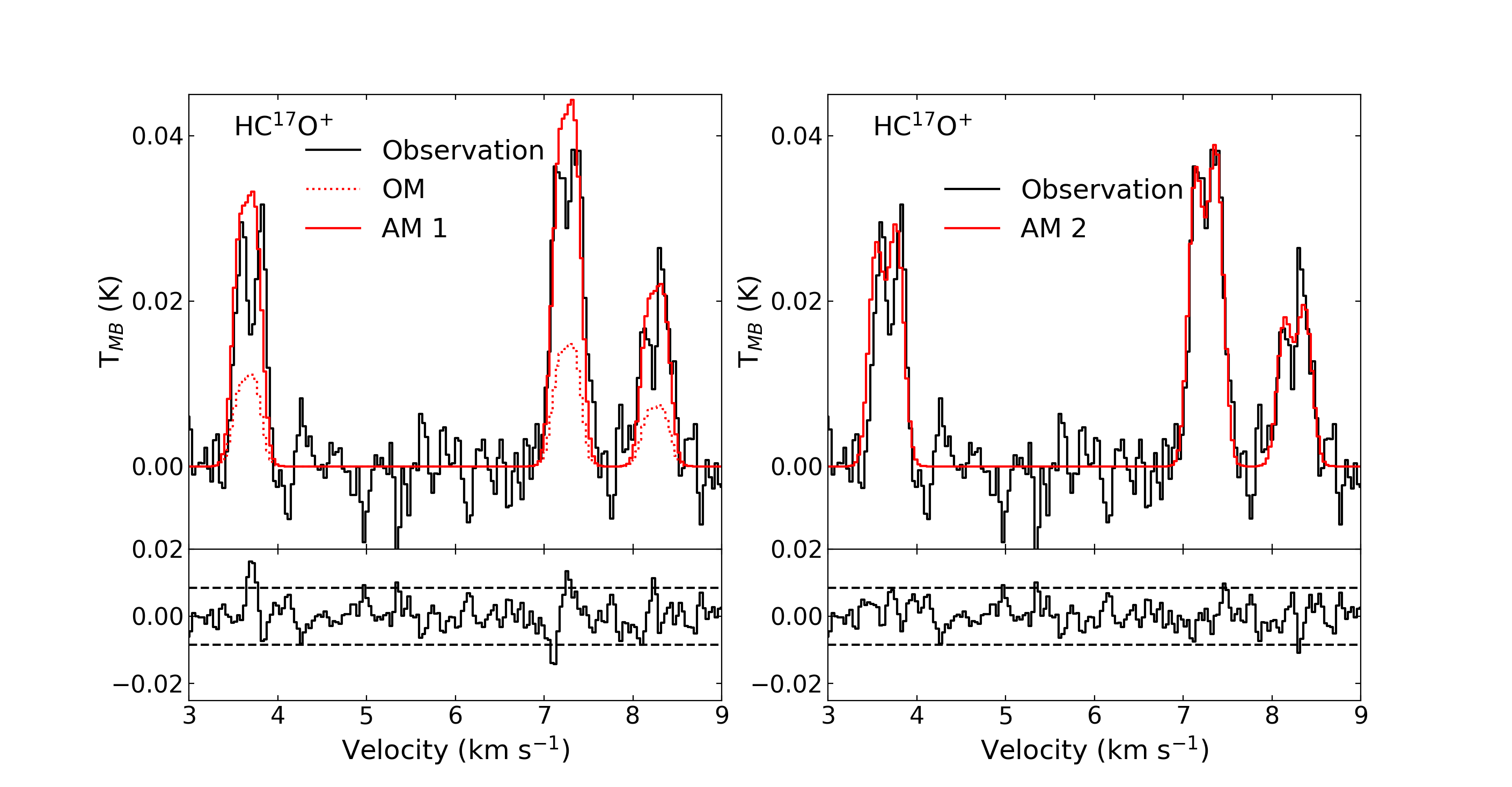}
\caption{\textit{Left.} Observed spectrum of the HC$^{17}$O$^{+}$(1-0) (black), the result of a simulation using the original \hcopp \, fractional abundance profile at $5\times10^{5}$ yrs and original pre-stellar core model (Original Model (OM), dotted red) and the result of a simulation using the \hcopp \, fractional abundance profile at $5\times10^{5}$\,yrs upscaled by a factor of 3 times that of the original pre-stellar core model (Adopted Model 1 (AM 1), solid red). Residuals are computed with the Adopted Model and are shown in the lower panel. The dashed lines represent the 3$\sigma$ levels. \textit{Right.} Observed spectrum of the HC$^{17}$O$^{+}$(1-0) (black) and the result of a simulation using the \hcopp \, fractional abundance profile at $5\times10^{5}$ yrs upscaled by a factor of 3 and an upscaled velocity profile by 30\% (Adopted Model 2 (AM 2), solid red). Residuals are computed with the Adopted Model and are shown in the lower panel. The dashed lines represent the 3$\sigma$ levels.}
\label{22987}
\end{figure*}

The observed \hcopp(1-0) can be reproduced by using the pre-stellar core model and the \hcopp \, fractional abundance profile at $5\times10^{5}$\,yrs with small modifications. Results are shown in the two panels of Figure \ref{22987}. The top left panel shows the Original Model (OM) \textsc{loc} fit, and the Adopted Model 1 (AM 1) \textsc{loc} fit over the observations. The OM histogram is the spectrum \textsc{loc} produces when using the original pre-stellar core model (from \cite{keto:15}) and original \hcopp \, fractional abundance profile. The AM 1 histogram is the spectrum \textsc{loc} produces when using the original pre-stellar core model \cite{keto:15} and a scaled-up (by a factor of 4) \hcopp \, fractional abundance profile. We use AM 1 to compute the residuals, which can be found on the panel under the spectra. In order to reproduce the intensity of the hyperfine components, we had to scale up the \hcopp \, fractional abundance profile by a factor of 4. However, this model fails to reproduce the double-peak morphology of the lines and for this a further change is needed. 

The top right panel of Figure \ref{22987} shows, with solid red histogram, a different Adopted Model 2 (AM 2) \textsc{loc} fit overlaid on the observations in black. The AM 2 corresponds to AM 1 with a scaled-up (by 30\%, ranging from -0.4 at the velocity peak to -0.01 km s$^{-1}$ at the edge of the core) velocity profile instead of the velocity profile from the pre-stellar core physical model of \cite{keto:15}. We use the AM 2 to compute the residuals, which can be found on the panel under the spectra. There is a reduction of the residuals standard deviation by 0.7 mK when compared to the AM 1 in the left panel, which only takes into account the \hcopp \, fractional abundance profile scaling. The discussion on the reasons behind the adjustments of the original pre-stellar core model is discussed in Section \ref{discussion}.\  
\raggedbottom
\vspace{0.25cm}

\begin{figure}[H]
\includegraphics[width=\columnwidth]{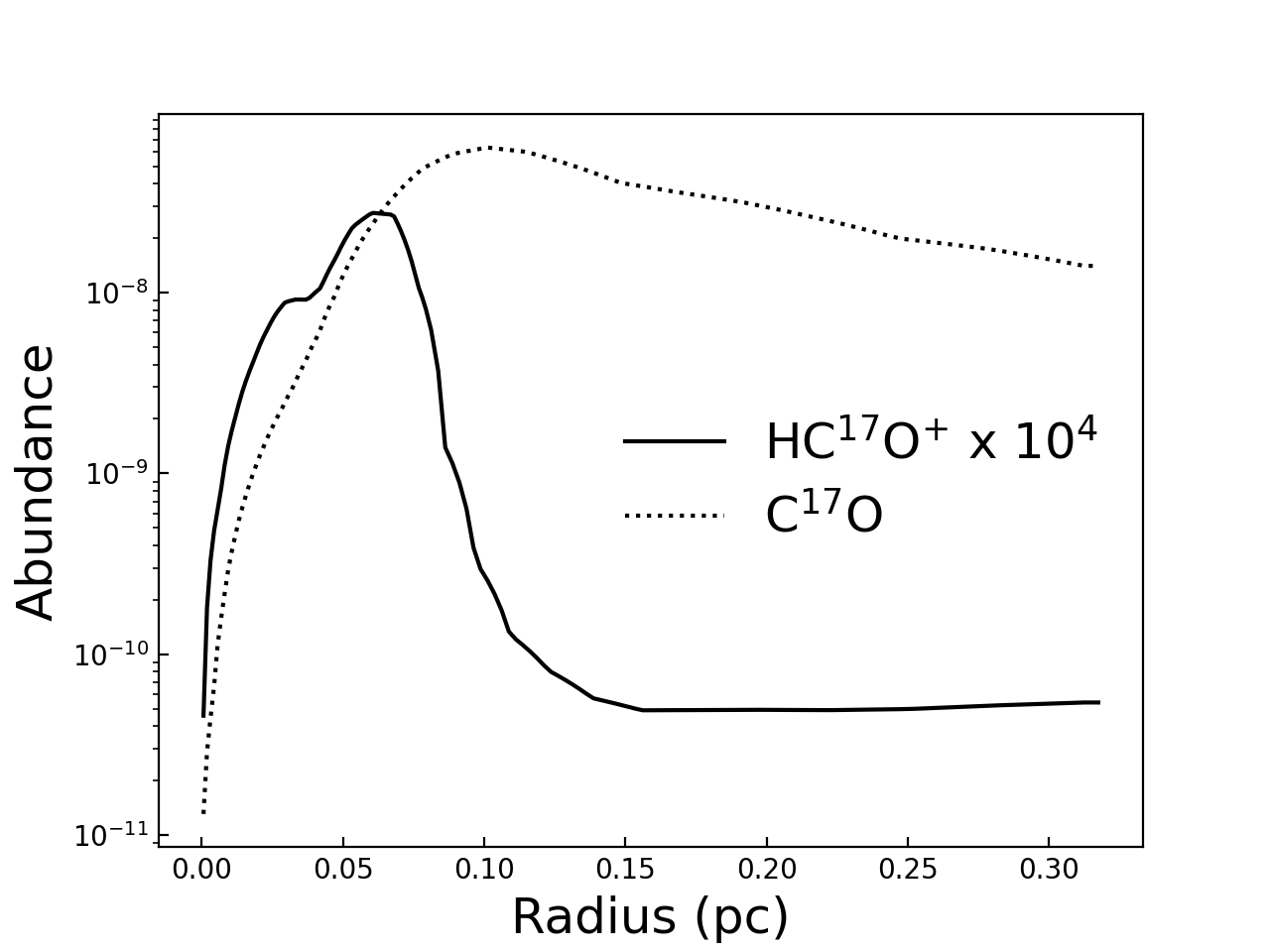}
\caption{The dotted curve shows the C$^{17}$O fractional abundance (abundance with respect to H$_{2}$) as a function of radius. The solid curve shows the fractional abundance of HC$^{17}$O$^{+}$, scaled by 10$^{4}$ computed with the chemical code from \cite{sipila:15, sipila:19} by scaling the CO and HCO$^{+}$ fractional abundance profiles by the isotope ratio [$^{16}$O/$^{17}$O] = 2044 \citep{penzias:81, wilson:94}.}
\label{abun}
\end{figure}

\begin{figure}[h]
\centering
\includegraphics[width=8.5cm]{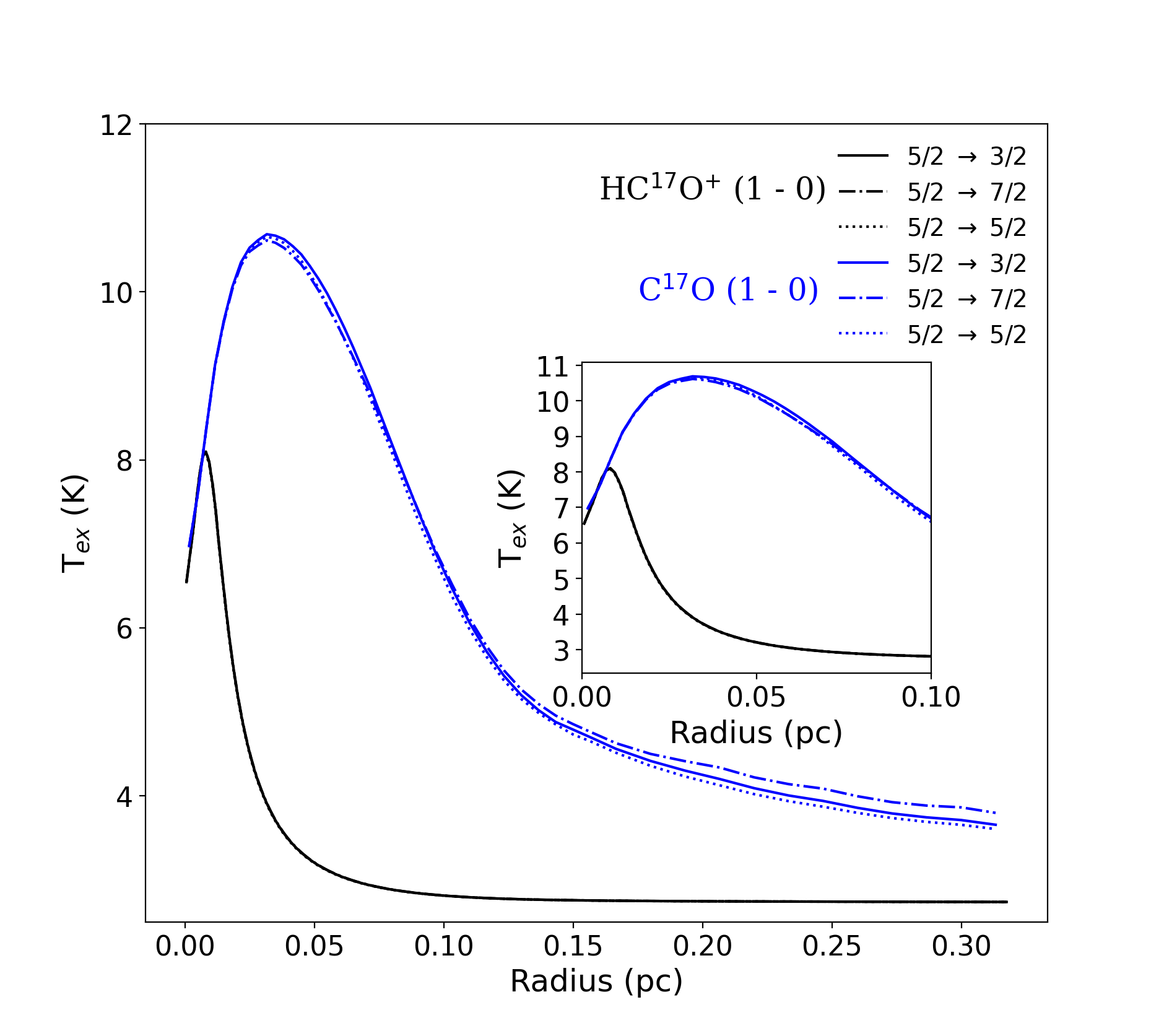}
\caption{Excitation temperature profile of the three \hcopp(1-0) and C$^{17}$O(1-0) hyperfine transitions across the core. The \hcopp(1-0) 5/2 $\rightarrow$ 3/2, 5/2 $\rightarrow$ 7/2 and 5/2 $\rightarrow$ 5/2 transitions are plotted with a black solid line, a black dash-dotted line, and with a dotted black line, respectively. The profiles for \hcopp \, are superimposed. The C$^{17}$O(1-0) 5/2 $\rightarrow$ 3/2, 5/2 $\rightarrow$ 7/2 and 5/2 $\rightarrow$ 5/2 transitions are plotted with a blue solid line, a blue dash and dotted line, and with a dotted blue line, respectively. The inset presents a zoom-in of the radial $T_\mathrm{ex}$ profile from 0 to 0.1 pc. }
\label{tex}
\end{figure}

Using the \hcopp \, fractional abundance profile in the Adopted models of Fig.\,\ref{22987} and the physical structure given in input to \textsc{loc},  we can derive the column density of \hcopp. This is done by integrating the multiplication of the fractional abundance profile (see Figure \ref{abun}) by the gas density and convolving to the beam size (28"). The column density derived from the fractional abundance profile used to model the observations in Figure \ref{22987} is 5.4$^{+0.7}_{-0.9}\times10^{10}$ cm$^{-2}$. The column density uncertainty is derived with the approach used in \cite{redaelli:18}. We compute several models from scaled fractional abundance profiles, from which we derive corresponding column density values. $\chi^2$ values are derived from the comparison of these models with the observations. We then plot $\chi^2$ versus the column density values. The column density uncertainties are derived setting a $\chi^2$ upper/lower limits to 15\%. The excitation temperature computed by \textsc{loc} for the three \hcopp(1-0) hyperfine transitions throughout the core is plotted in Figure \ref{tex} in black. We can see that the $T_\mathrm{ex}$ profiles of the different hyperfine components do not vary significantly. In order to compare this method with the column density calculation from the line fitting with Gaussians in a C$T_\mathrm{ex}$ (constant $T_\mathrm{ex}$) approximation (see \citealt{caselli:02}), we perform HyperFine Structure (HFS) fits in \textsc{class}. Next, assuming an excitation temperature of 5 K, we calculated a \hcopp \, column density of 4.1$\pm0.3\times10^{10}$ cm$^{-2}$. The choice for the $T_\mathrm{ex}$ comes from N$_{2}$H$^{+}$(1-0), which has similar critical density to \hcopp(1-0) (1.4$\times10^{5}$ and 1.5$\times10^{5}$ cm$^{-3}$ respectively), observed towards L1544 in \cite{crapsi:05}. For this calculation we use the optically thin approximation as we have confirmed the \hcopp \, $J$ = 1 - 0 transition to be optically thin (details of the calculation can be found in Appendix \ref{ctex}). The column density derived with the simple C$T_\mathrm{ex}$ method is in agreement within errors with the one derived with our model. Furthermore, we compare these two column density values with the one obtained by scaling the HC$^{18}$O$^{+}$ column density derived in \cite{redaelli:19} by [$^{18}$O/$^{17}$O]= 3.67, yielding a \hcopp \, column density of 4.6$\pm0.3\times10^{10}$ cm$^{-2}$.\ 

\begin{figure*}[t]
\centering
\includegraphics[width=\textwidth]{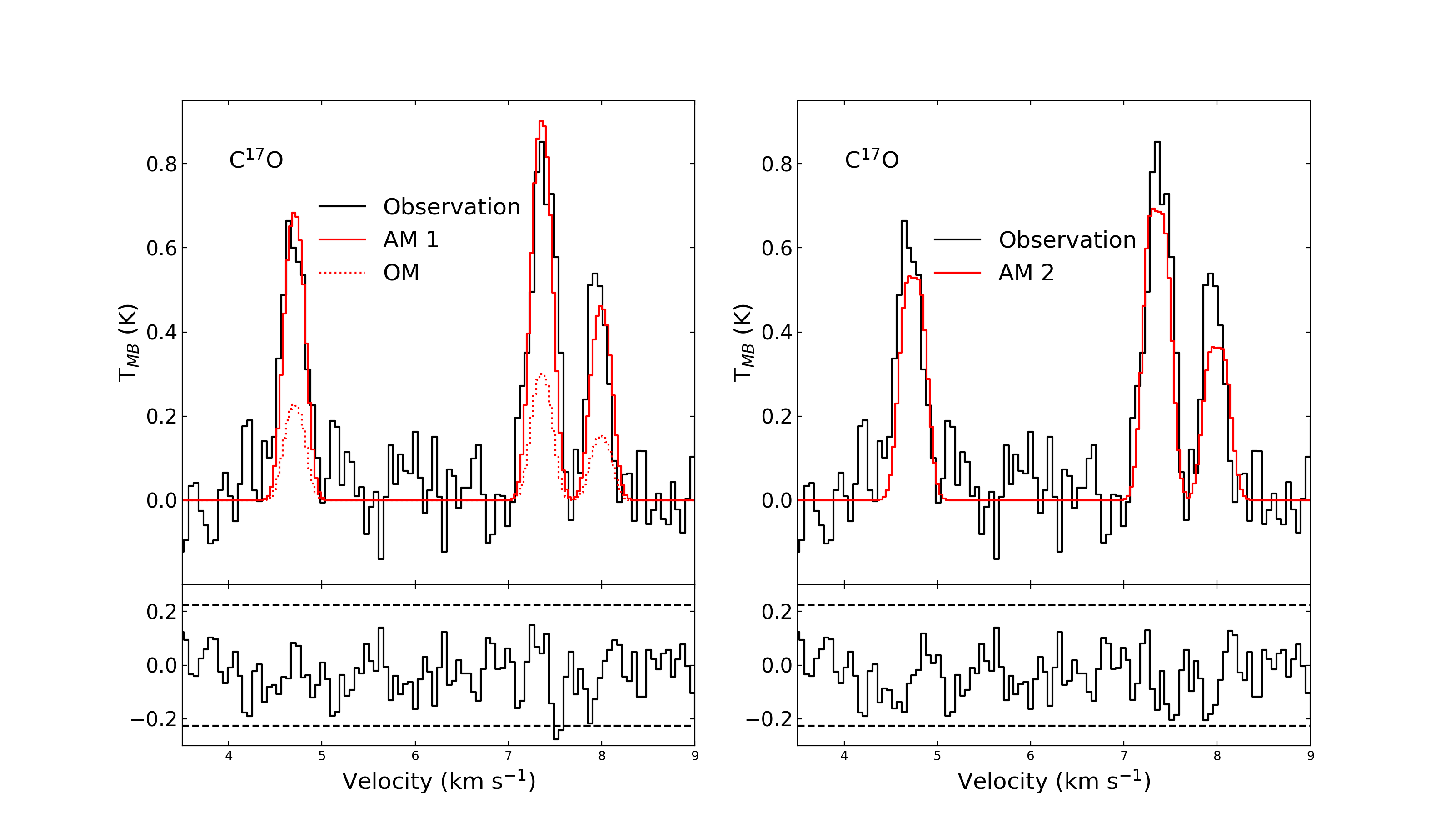}
\caption{\textit{Left.} Observed spectrum of the C$^{17}$O (1-0) (black), the result of a simulation using the original C$^{17}$O fractional abundance profile at $5\times10^{5}$\,yrs and the original pre-stellar core model (Original Model (OM), dotted red) and the result of a simulation using the C$^{17}$O fractional abundance profile at $5\times10^{5}$\,yrs upscaled by a factor of 3 and the original pre-stellar core model (Adopted Model 1 (AM 1), solid red). Residuals are computed with the Adopted Model and are shown in the lower panel. The dashed lines represent the 3$\sigma$ levels. \textit{Right.} Observed spectrum of the C$^{17}$O (1-0) (black) and the result of a simulation using the C$^{17}$O fractional abundance profile at $5\times10^{5}$\,yrs increased by a factor of 3 and a scaled-up velocity profile by 30\% (Adopted Model 2 (AM 2), solid red). Residuals are computed with the Adopted Model and are shown in the lower panel. The dashed lines represent the 3$\sigma$ levels.}
\label{c17or}
\end{figure*}

\subsection{\texorpdfstring{C$^{17}$O}{C17O}} \label{c17o}

The $J$ = 1 - 0 transition of C$^{17}$O has been observed towards L1544 by \cite{chacon:19}. Contrary to \hcopp, C$^{17}$O presents single-peaked hyperfine components, despite also being an optically thin transition. The critical densities of \hcopp(1-0) and C$^{17}$O(1-0) are $1.5\times10^{5}$ cm$^{-3}$ and $2.0\times10^{3}$ cm$^{-3}$, respectively, as calculated from the Einstein and collisional coefficients at 10 K from the Leiden Atomic and Molecular Database (LAMDA)\footnote{https://home.strw.leidenuniv.nl/$\sim$moldata/}. To model the hyperfine C$^{17}$O emission we required the hyperfine collisional rate coefficients which we approximated from the non-hyperfine collisional rate coefficients available at LAMDA \textbf{\cite{yang:10}} with a $Mj$ randomisation approach (Appendix \ref{c17ocoll}). We note that \cite{dagdigian:22} has recently also computed hyperfine C$^{17}$O collisional rate-coefficients using the recoupling technique. Like in the case of \hcopp, we are able to reproduce the observed spectrum with the pre-stellar core model detailed in Section \ref{loc}, and the fractional abundance profile at $5\times10^{5}$ yrs described in Section \ref{chemical}, with a scaling factor of 3. Results are shown in the two panels of Figure \ref{c17or}. The top left panel shows the Original Model (OM) \textsc{loc} fit, and the Adopted Model 1 (AM 1) \textsc{loc} fit over the observations. The OM histogram is the spectrum \textsc{loc} produces with the original pre-stellar core model and original C$^{17}$O fractional abundance profile. The AM 1 histogram is the spectrum \textsc{loc} produces with the original pre-stellar core model and an fractional abundance profile of C$^{17}$O scaled-up by a factor of 3, needed to fit the intensity of the observed lines (similar to what found in the case of \hcopp). The AM 1 is the model we use to compute the residuals, which can be found on the panel under the spectra. They show an overall good agreement between the AM 1 and the observations. 

The top right panel of Figure \ref{c17or} shows the Adopted Model 2 (AM 2) \textsc{loc} spectrum overlaid on the observations. The AM 2 corresponds to the AM 1 with a velocity profile scaled-up by 30\% (ranging from -0.4 at the velocity peak to -0.01 km s$^{-1}$ at the edge of the core). We use the AM 2 to compute the residuals, which can be found on the panel under the spectra. The right bottom panel shows the residuals obtained by subtracting the observed spectrum from the AM 2 spectrum. Contrarily to \hcopp, there is an increase of the residuals standard deviation by 1.5 mK when a scaled-up velocity profile is adopted. Therefore, the best model for reproducing the C$^{17}$O observed spectrum is the one that uses the increased (by a factor of 3) C$^{17}$O fractional abundance profile and the original pre-stellar core velocity profile (AM 1 in upper left panel of Figure \ref{c17or}). The reason behind this difference between \hcopp (1-0) and C$^{17}$O(1-0) observations is discussed in Section \ref{discussion}.\ 

We have also computed the C$^{17}$O column density from the fractional abundance profile used in \textsc{loc} to fit the observations as done with \hcopp. The excitation temperature computed by \textsc{loc} for the three C$^{17}$O(1-0) hyperfine transitions throughout the core is plotted in Figure \ref{tex} in blue. The $T_\mathrm{ex}$ profiles of the different hyperfine components do not vary significantly from each other. The calculated C$^{17}$O column density is $4.8
^{+0.7}_{-0.8}\times10^{14}$ cm$^{-2}$ (the uncertainty is calculated in the same way as for \hcopp). The column density calculated in \cite{chacon:19} is $6.8\pm0.6\times10^{14}$ cm$^{-2}$, 41\% higher than the column density extracted from the C$^{17}$O fractional abundance profile used for the modelling with \textsc{loc}. This difference in column densities could be due to the $T_\mathrm{ex}$ used; \cite{chacon:19} assume a constant $T_\mathrm{ex}$ of 10 K across the core, while \textsc{loc} computes the column density using the $T_\mathrm{ex}$ radial profile (Figure \ref{tex}).

\section{Discussion} \label{discussion}

In past works, two scenarios have been discussed to explain the double-peaked morphology in optically thin transitions observed towards L1544 \citep{tafalla:98, caselli:02}: the presence of two velocity components in the line of sight and the depletion of the molecules towards the centre of the core.\

The double-peak structure of the optically thin C$^{34}$S (2-1) line is suggested to arise from two separated velocity components in the line of sight at 7.10 and 7.25 km s$^{-1}$ which overlap in the centre \citep{tafalla:98}. We discard the two velocity component scenario for our case, as this would produce double-peaked lines also from C$^{17}$O(1-0), in disagreement with observations, as shown in Fig.\,\ref{c17or}. Moreover, we are able to reproduce the \hcopp (1-0) line profile using a simple model of an individual object in contraction, as also found in the past using other molecular lines.\

In \cite{caselli:02}, it is suggested that the optically thin double-peaked line profiles of the $F_{1}, F$ = 1,0 $\rightarrow$  1,1 line of N$_{2}$H$^{+}$ (1-0) and HC$^{18}$O$^{+}$ (1-0) towards L1544 arises from the depletion of these molecules towards the centre of the core. To mimic this depletion the fractional abundance of N$_{2}$H$^{+}$ and HC$^{18}$O$^{+}$ was set to 0 between the centre and 2000 au and 1400 au, respectively. The size of these gaps in the fractional abundance profiles are related to the radius from the centre of the core at which these molecules present significant depletion. This artificial depletion reproduced the $F_{1}, F$ = 1,0 $\rightarrow$  1,1 line of N$_{2}$H$^{+}$ (1-0) and HC$^{18}$O$^{+}$ (1-0) line profiles. To test whether an enhanced depletion of HC$^{17}$O$^{+}$ towards the core centre would reproduce the observations, we created a "hole" in the fractional abundance profile similarly to \cite{caselli:02} (more information in Appendix \ref{dep}), but the central depletion of \hcopp \, does not reproduce the observed line profile (e.g. Figure \ref{20666}). This was expected as the chemical model already naturally predicts the depletion of HCO$^{+}$ towards the central part of the core.\ 

Self-absorption is not expected to produce double peaks in lines of low  species, as their concentration in the low density foreground layer is not large enough to absorb photons emitted in the dense core. Nevertheless, self-absorption has been considered as a possible contribution to double-peaked line morphologies for some transitions in past works \citep{caselli:99, caselli:02}. To make sure that a sufficient  of \hcopp \, in a foreground layer would not induce self absorption, we modelled the observations by adding an  profile from 0.3 to 1.5 pc with a \hcopp \, layer corresponding to a visual extinction of A$_{v}$=4 mag (more information in the Appendix \ref{sa}). Also this test did not reproduce the observed spectrum, as shown in Figure \ref{22180}.\ 

One of the parameters that improved the fitting of lines in L1544 in past works is the velocity profile.  In \cite{bizzocchi:13}, upscaling the velocity profile by 75\% was necessary to reproduce the high sensitivity observations of N$_{2}$H$^{+}$(1-0)  towards L1544. Both the observed linewidth and line profiles are well reproduced with \textsc{loc} using a constant  alongside the upscaled velocity profile from the \cite{keto:08} physical structure model (Figure 5 in \cite{bizzocchi:13}). The upscaling of the velocity profile allows the further splitting in velocity of the two contracting parts of the cloud laying in front and behind the core centre where the density is close to the critical density of the transition ($1.5\times10^{5}$ cm$^{-3}$). As in \cite{bizzocchi:13}, in our case this is reflected in the split of the \hcopp(1-0) hyperfine components in agreement with the observed spectrum shown in Figure \ref{hc17o+}.\ 

Infall motions were also invoked to explain the double-peaked profiles of HC$^{18}$O$^{+}$ 1-0 and N$_{2}$H$^{+}$ 1-0 towards the dust peak of L1544 in \cite{vandertak:05}. HC$^{18}$O$^{+}$ presents a line split where the two peaks appear with the same intensity, while N$_{2}$H$^{+}$ presents the well-known blue-asymmetry in the peak intensity indicating that the line is self-absorbed.  In this paper they combine the core physical model from \cite{galli:02} with infall models at "t3" and "t5" time steps which correspond to 2.660 and 2.684 Myr after the start of the core collapse, respectively from \cite{ciolek:00}. Nevertheless, this model was not able to reproduce the H$_{2}$D$^{+}$ 1$_{10}$-1$_{11}$spectral line profile. In our work, we used a physical core model \citep{keto:101} which has been tailored specifically for L1544, and includes infall dynamics which are coherent with the physical structure, as well as a state-of-the-art chemical model \citep{sipila:15}. Our model has reproduced a great number of molecular lines (see Section \ref{loc}) including HC$^{18}$O$^{+}$ (1-0) and N$_{2}$H$^{+}$ (1-0) in \cite{redaelli:19}. Moreover, the higher spectral resolution (by 0.02 km s$^{-1}$) of our \hcopp \, (1-0) spectrum and the hyperfine splitting allows us to refine the physical model by fitting the line profile, compared to previous HC$^{18}$O$^{+}$ (1-0) observations. The velocity profiles used in \cite{vandertak:05} range, for "t3" and "t5", respectively, from -0.01 to -0.14 and -0.01 to -0.19 km s$^{-1}$ (see Figure 2 in \cite{ciolek:00}). These velocity profiles differ from the one we have used to model \hcopp, where the velocity peak has an average of 2.5 times higher velocity than the maximum velocity in the profiles used for the modelling in \cite{vandertak:05}.\

The different line profiles of \hcopp(1-0) and C$^{17}$O(1-0) arise from their different critical densities and spatial distributions in the core. \hcopp \, and C$^{17}$O have different critical densities ($1.5\times10^{5}$ cm$^{-3}$ and $2.0\times10^{3}$ cm$^{-3}$, respectively). The  profile and critical density combination makes the \hcopp \, line to emit preferentially closer to the core centre (see its excitation temperature profile in Fig.\,\ref{tex} in black), while C$^{17}$O has a more extended emission (see its excitation temperature profile in Fig.\,\ref{tex} in blue). As \hcopp \, emits in a small region close to the centre, where the contraction velocity also has a peak, the spectrum results in a double peak (where the local contraction velocity is 30\% higher than in the original model of \cite{keto:15}). Unlike \hcopp, C$^{17}$O emission arises from lower density gas further away from the core centre, where the contraction velocity is lower (see Fig.\,\ref{pm}), resulting in a single peak. The fact that an upscaled velocity profile fits the \hcopp(1-0) line profile but not C$^{17}$O simply tells us that the infall velocity is higher towards the inner part of the core, as expected in a contracting Bonnor Ebert sphere \citep{keto:10} at a more evolved stage compared to the one adopted by \citet{keto:15}. We estimate that it would take an additional $1\times10^{4}$ yrs for the infall velocity peak to increase by 30\%. 
 We must also note that L1544 is not spherically symmetric, as assumed so far, but it has an elongated structure \citep{caselli:02}. More complex simulations have been done taking into account a flattened structure for L1544 which is in closer resemblance to reality \citep{caselli:19}. Nevertheless, \cite{caselli:22}, have shown that the density and velocity profiles along the major and minor axes of the 3D simulation are very close to the same profiles in the spherically symmetric model of \cite{keto:14},  reinforcing the accuracy of the \citet{keto:15} model despite its simplicity.\ 
The need to take into account the physical structure and kinematics of a source when interpreting its spectra is made evident in this work. Using a slab model that does not take into account the physical nor the kinematic structure of the core can lead to the derivation of non-accurate parameters and, in this case, not reaching any conclusions about the nature of the physical process behind the double-peaked profiles.\
As we have seen from Section \ref{results}, both \hcopp \, and C$^{17}$O models required an upscaling of the fractional abundance profile used to fit the observations. The necessary \hcopp \, fractional abundance adjustment required to fit the observations is done by adjusting the HCO$^{+}$ profile predicted by our chemical code (see Section \ref{chemical}). We note that the HCO$^{+}$ fractional abundance profile resulting from a static model, where the physical structure is kept fixed while the chemistry is evolving, differs from the one computed with a dynamic model, as can be seen in the central-upper panel in Figure 12 in \cite{sipila:18}. The fractional abundance profile of HCO$^{+}$ in the static model appears $\sim$3 times higher than the profile obtained with the dynamic model at the very centre of the core (until $2.0\times10^{4}$ au) and $\sim$4 times lower from $2.0\times10^{4}$ to $2.5\times10^{4}$ au. However, the physical model discussed in \cite{sipila:18} refers to a dense core with a mass of 7.2 \(\textup{M}_\odot\) which is significantly lower than that of L1544 described by \cite{keto:101} (10 \(\textup{M}_\odot\)), so the dynamic model results cannot be used here. We are presently preparing a new dynamical model for L1544, which will help to constrain the issue of static vs. dynamic models (Sipil\"a et al., subm.). Finally, the fact that both the fractional abundances of \hcopp \, and C$^{17}$O needed to be upscaled by a similar factor can be the consequence of too large CO depletion compared to the observations, also suggesting that dynamics could help in better reproducing our data (as less CO-depleted material keep moving toward the central regions, unlike in the static model). \

\section{Conclusions}\label{conclusions}

In this article we have presented new, high-sensitivity and high-resolution \hcopp \, $J$ = 1 - 0  observations towards the dust peak of L1544, which unveiled the double-peaked nature of its hyperfine components. By carrying out a full non-LTE radiative transfer modelling of \hcopp(1-0) with new hyperfine collisional-rate coefficients, we have explored what is causing the double-peak profile observed in optically thin transitions of high density tracers. The power of fully taking into account radiative transfer is made evident in this work where a modelling of the entire source is needed to reproduce the observed line profile. Moreover, we have tested the model by reproducing previous observations of the C$^{17}$O(1-0) line which presents a single-peak line morphology. Our main conclusions are as follows:

\begin{itemize}

    \item The \hcopp(1-0) line profile can be reproduced with a model of a contracting pre-stellar core, finding no evidence for separate velocity components along the line of sight. This indicates that a double-peaked profile towards a centrally condensed core can be solely an effect of radiative transfer in a contracting centrally concentrated dense core. In the present work the reason for the double peak lies in the infall velocity and the critical density of the line. This makes the line sensitive to the motions at radii of about 0.01\,pc where the volume density is close to the critical density, and the contraction velocity has a peak.
    
    \item The fact that an upscaled velocity profile was required to reproduce \hcopp \, but not C$^{17}$O suggests that the L1544 pre-stellar core is more dynamically evolved than previously thought (1$\times10^{4}$ yrs), as higher (by 30\%) velocities, compared to the original physical model, are required toward the core centre to reproduce the splitting. Higher central velocities are in fact expected in contracting BE spheres at later stages of evolution \citep{keto:101}. 
    
    \item Both the \hcopp \, and C$^{17}$O fractional fractional abundances are underproduced by our chemical models by a factor of 4 and 3 respectively. This suggests that predicted CO freeze-out is too large, possibly due to the physical model not evolving dynamically.

\end{itemize}

\textit{Acknowledgements.} J.F.A., S.S., P.C., F.O.A., O.S, E.R. and L.B gratefully acknowledge the support of the Max Planck Society.

\bibliographystyle{aa}
\bibliography{vanderwaals}

\appendix 

\section{Considered scenarios results}

In this section we present complementary results obtained to evaluate the viability of the different possible scenarios for the \hcopp(1-0) double-peaked structure. All the models presented have been computed with a HCO$^{+}$ fractional abundance profile at $t$ = $5\times10^{5}$ yrs so they are comparable with the best fit results (AM 2 in Figure \ref{22987}). Fractional abundance profiles at time steps between 10$^{4}$ and 10$^{7}$ yrs were also tested leading to less accurate fits with larger residuals standard deviation.\ 

\subsection{Depletion}\label{dep}

The chemical model takes into account the CO depletion by freeze-out at the centre of the core which affects the HCO$^{+}$ fractional abundance (Figure \ref{abun}), as this molecule is formed by the reaction of CO with H$_{3}^{+}$.\ 

\begin{figure}[H]
\centering
\includegraphics[width=9cm]{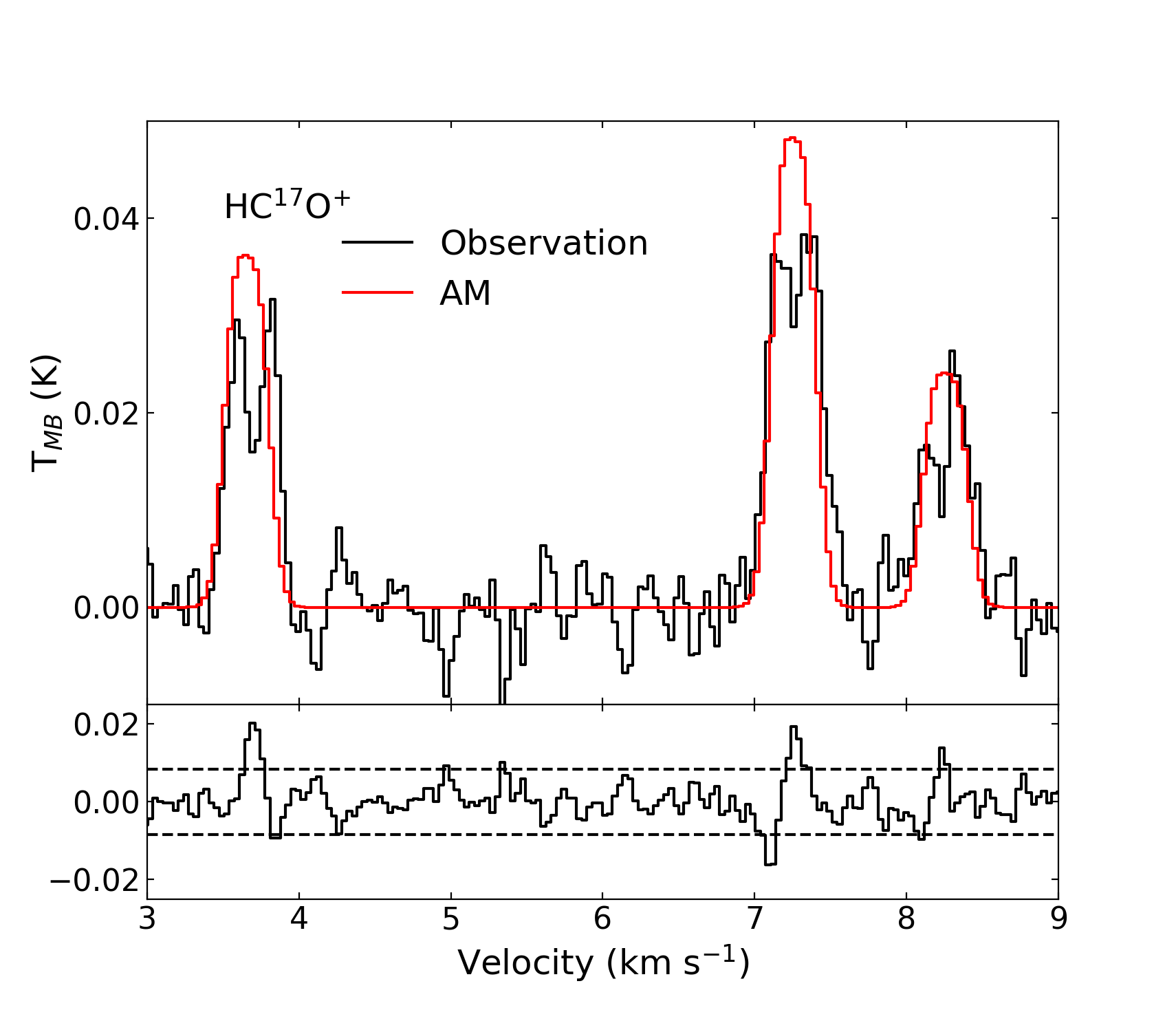}
\caption{Spectrum of the HC$^{17}$O$^{+}$ (1-0) observation (black) and product of model with the upscaled by a factor of 16 fractional abundance profile at $5\times10^{5}$ yrs and modified to have 0 fractional abundance from 0 to 7000 au (red).}
\label{20666}
\end{figure}

Following the idea that the model flat-topped profiles in the OM of Figure \ref{22987} could be the result of the enhanced \hcopp \, depletion towards the core centre, we test larger depletions than the ones taken into account in the chemical model, to see whether we can fit the observed line profiles better. For that, we model the radiative transfer of the molecule with a modified fractional abundance profile creating fractional abundance "holes" with different radii. We test the effect of a 7000 au radius "hole" and show the resulting spectra from a model with \hcopp \, depletion from 0 to 7000 au (Figure \ref{20666}). Notice that, in order to be able to distinguish the feature of the lines properly, the \hcopp \, fractional abundance had to be increased by a factor of 16 to make the spectrum visible in the figure. We obtain a faint line emission with a double-peak feature which is not enhanced from the OM profile in Figure \ref{22987}. Therefore, the \hcopp \, depletion towards the centre of the core alone cannot explain the observed double-peaked structure.\

\subsection{Self-absorption}\label{sa}

\begin{figure}[H]
\centering
\includegraphics[width=9cm]{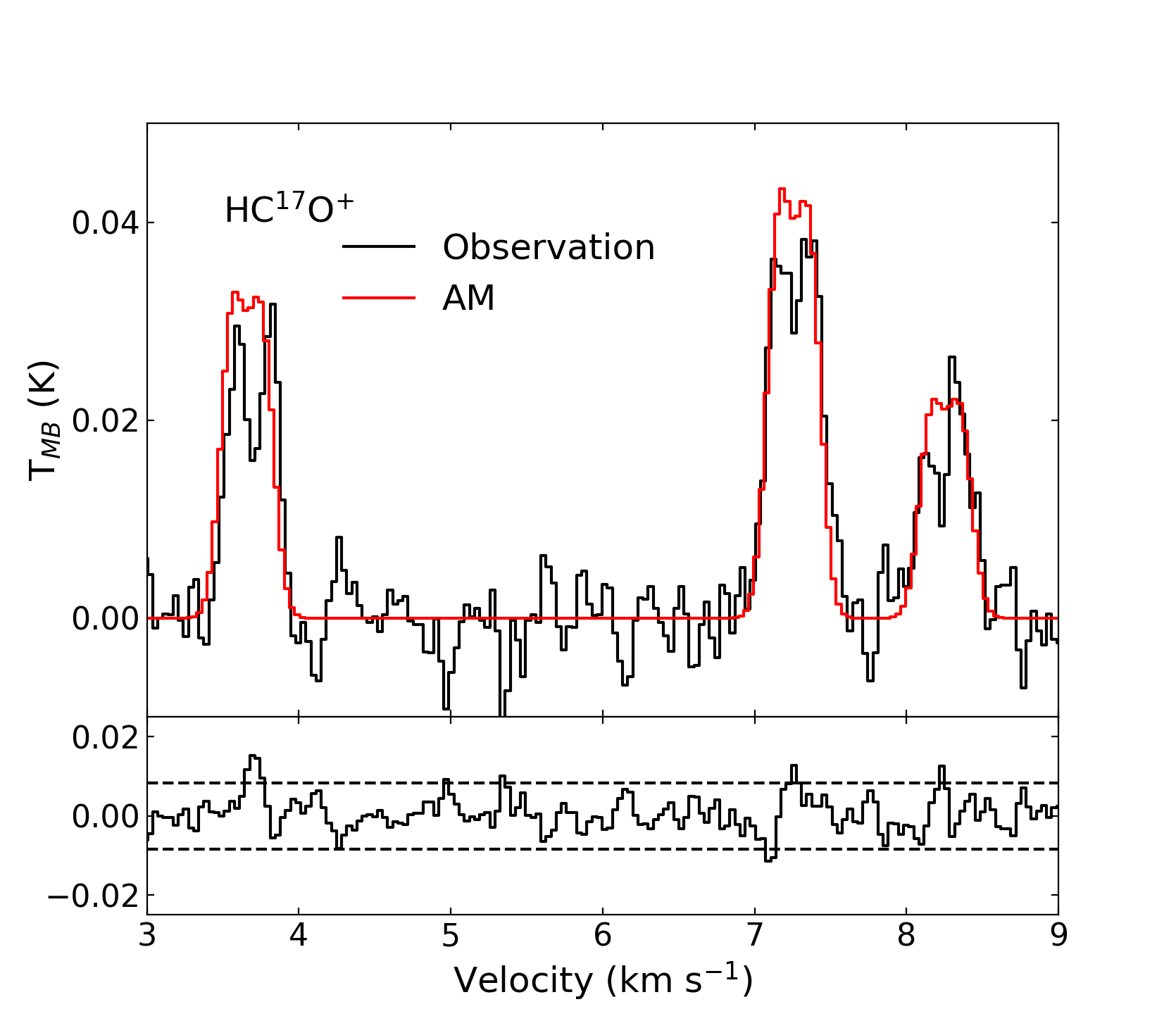}
\caption{Spectrum of the HC$^{17}$O$^{+}$ (1-0) observation (black) and product of model with an upscaled by a factor of 3 extended fractional abundance profile with A$_{v}$=4 at $5\times10^{5}$ yrs.}
\label{22180}
\end{figure}

Self-absorption is not expected for an optically thin transition such as \hcopp(1-0). Nevertheless, we tested the effects of an envelope-like extension added to the fractional abundance profile corresponding to a visual extinction of $A_\mathrm{v}$=4 mag which lengthens the fractional abundance profile from 0.3 to 1.7 pc. In Figure \ref{22180} we present the $A_\mathrm{v}$=4\,mag extended-model spectrum with a \hcopp \, fractional abundance profile scaled up by a factor of 3 to be able to distinguish possible self-absorption features in the line profiles. This trial does not show signs of \hcopp(1-0) self-absorption nor reproduces the observed line profile which gives evidence that an enhanced \hcopp \, molecular fractional abundance in the foreground does not account for the observed line shape.\ 

\section{\texorpdfstring{Comparison with Constant $T_\mathrm{ex}$ (C$T_\mathrm{ex}$) approximation}{Comparison with Constant Tex (CTex) approximation}} \label{ctex}

To derive the column density from the observations we first check that the \hcopp(1-0) transition is optically thin. A first fit with \textsc{class} returns larger optical depth uncertainties than optical depth values. In the optically thin approximation ($\tau_{\nu}$<0.1; where $\tau_{\nu}$ represents the total optical depth) $\tau_{\nu}$ is not included in the radiative transfer equation and thus the problem degenerates resulting in the large optical depth uncertainties observed. As the optical depth derivation for this case with \textsc{class} is not possible, we estimate its value with the following equation, assuming three different $T_\mathrm{ex}$ values (5, 7 and 9 K): 

\begin{equation}
\label{tau}
  \tau_\nu = -ln [1-\frac{T_\mathrm{MB}}{[J_{\nu}(T_{ex})-J_{\nu}(T_{bg})]}] .
\end{equation}

\noindent
The total optical depth is represented by $\tau_\nu$, the line brightness temperature by $T_\mathrm{MB}$ and $J_{\nu}(T_{ex})$ and $J_{\nu}(T_{bg})$ correspond to the Rayleigh-Jeans Equivalent Temperature at $T_{ex}$ and $T_{bg}$ respectively. The optical depth obtained for the line lies between 0.03 and 0.08 which confirms that the $J$ = 1 - 0 rotational transition of \hcopp \, is optically thin.

We then proceed to fit the observations with the Hyperfine Structure (HFS) tool in \textsc{class}. This tool fits the hyperfine components of a transition using laboratory calculated frequencies, assuming a Gaussian velocity distribution, the same excitation temperature of the hyperfine components and LSR velocity. In our particular case, we also fixed the optical depth to the minimum value allowed (0.1) to indicate the line is optically thin. In order to reproduce the double peak, as kinematic information cannot be included in \textsc{class}, we apply this tool taking into account two separate velocity components. This fit results in a hyperfine component model that follows the statistical intensity ratios. In non-C$T_\mathrm{ex}$ conditions, the relative intensity amongst hyperfine components can differ from the statistical value. In our case, however, the ratio of intensities does not differ significantly from the statistical values.\

\begin{figure}[H]
\centering
\includegraphics[width=8cm]{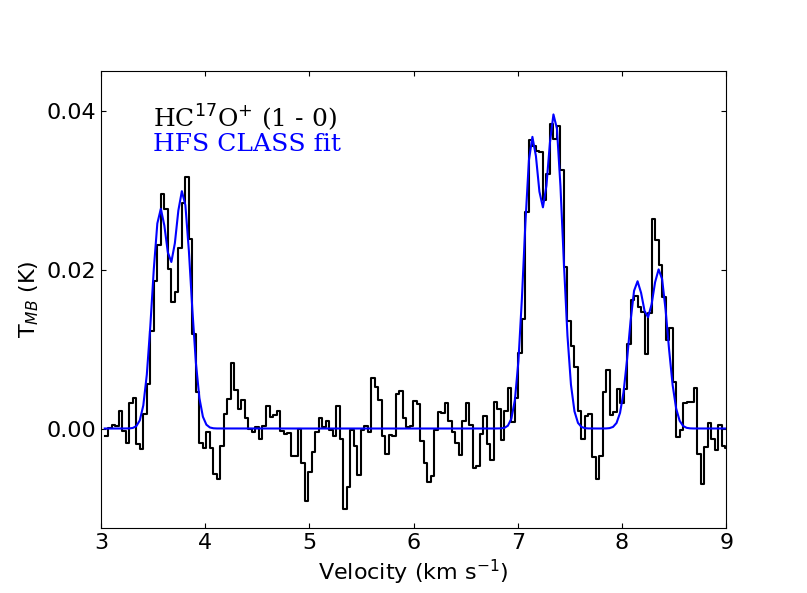}
\caption{Spectrum of the \hcopp \, $J$ = 1 - 0 at the dust peak of L1544 (black). HFS fit done with CLASS (blue).}
\label{hfs}
\end{figure}

\begin{table}[H]
\begin{center}
\caption{Results of the \textsc{class} HFS fit of the observed \hcopp(1-0) rotational transition towards L1544, treating the two peaks of each hyperfine line as separate velocity components.}
\begin{tabular}{ ccccc } 
\hline\hline
 v$_{LSR}$ & $T_\mathrm{MB}$ & $\Delta$v\\
 (km s$^{-1}$) & (K) & (km s$^{-1}$)\\
\hline
 7.13$\pm$0.07 & 8.2$\pm0.4\times10^{-2}$ &  0.177$\pm$0.018\\
 7.35$\pm$0.07 & 9.0$\pm0.4\times10^{-2}$ &  0.186$\pm$0.017\\
 \hline
\end{tabular}
\label{param}
\end{center}
\end{table}

With the parameters derived from HFS fittings (Table \ref{param}) and the \hcopp \, molecular constants extracted from The Cologne Database for Molecular Spectroscopy (CDMS)\footnote{https://cdms.astro.uni-koeln.de}, we calculate the column density of the molecule assuming that the transition is optically thin. For this purpose, we use Equation \ref{n} \citep{mangum:15}: 

\begin{equation}
\label{n}
  N_\mathrm{TOT} = \frac{8\pi k_B \nu^2 W}{A_{ul}hc^3}\frac{Q_{rot}}{g_{u}} \frac{J_{\nu}(T_{ex})}{J_{\nu}(T_{ex})-J_{\nu}(T_{bg})} \exp{(\frac{E_u}{k_BT_{ex}})},
\end{equation}

\noindent
where $N_\mathrm{TOT}$ is the total column density, $k_\mathrm{B}$ is the Boltzmann constant, $\nu$ is the transition frequency, $A_\mathrm{ul}$ is the Einstein coefficient, \textit{h} is the Planck constant, \textit{c} is the light speed, $E_\mathrm{u}$ is the upper level energy and $W$ is the integrated intensity. As seen on Table \ref{param}, the HFS fit returns for each velocity component the summed $T_\mathrm{MB}$ of the three hyperfine lines and the average linewidth $\Delta$v. We calculate a global $W$ for each velocity component from the $T_\mathrm{MB}$ and $\Delta$v assuming a Gaussian profile. On Equation \ref{n}, $g_\mathrm{u}$ represents the combined upper level degeneracy of the hyperfine transitions, which equals 18. Lastly, $Q_\mathrm{rot}$ represents the partition function. We used three different $Q_\mathrm{rot}$; 16.54, 22.23 and 27.94 for three different $T_\mathrm{ex}$: 5, 7 and 9 K, used for the calculations. The result is six column density values; corresponding to two velocity components and three assumed $T_\mathrm{ex}$ (Table \ref{column}). \  

The column density of the two fitted velocity components are then summed for each $T_\mathrm{ex}$ resulting in the calculated total \hcopp \, column density. The derived total column densities are the same for all $T_\mathrm{ex}$ within uncertainties. We conclude that the column densities do not strongly depend on the $T_\mathrm{ex}$ used in the range of values here considered. The derived column density of \hcopp \, at $T_\mathrm{ex}$= 5\,K towards L1544 is 4.1$\pm0.3\times10^{10}$\,cm$^{-2}$. On the other hand, the extracted column density from the \hcopp \, fractional abundance profile used to fit the observational spectrum with \textsc{loc} is 5.4$^{+0.7}_{-0.9}\times10^{10}$\,cm$^{-2}$, consistent within errors with the column density obtained with \textsc{class}. Contrary to a C$T_\mathrm{ex}$ approach where a single excitation temperature is used, the column density derivation from the fractional abundance profile integration takes into account the excitation temperature profile across the core (Figure \ref{tex}).\

\begin{table}[H]
\caption{Derived column densities from HFS fitting.}
\label{column}
\centering
\begin{tabular}{ c c c c c}
\hline\hline
 $V_{lsr}$ & $N_{col}$ ($T_{ex}$=5 K) & $N_{col}$ ($T_{ex}$=7 K) & $N_{col}$ ($T_{ex}$=9 K)  \\
(km s$^{-1}$) & (cm$^{-2}$) & (cm$^{-2}$) & (cm$^{-2}$) \\
\hline
 7.13$\pm$0.03 & 1.9$\pm0.2\times10^{10}$ & 1.7$\pm0.2\times10^{10}$ & 1.7$\pm0.2\times10^{10}$\\
 7.35$\pm$0.03 & 2.2$\pm0.2\times10^{10}$ & 1.9$\pm0.2\times10^{10}$ & 2.0$\pm0.2\times10^{10}$\\

\hline
\end{tabular}
\end{table}

\section{\hcopp Hyperfine Collisional Rate Coefficients}\label{colcoe}

In this section we present for the first time the \hcopp–H$_{2}$ hyperfine collisional rate coefficients calculated with the method described in Section \ref{hyper}. For each transition $J$, $F \rightarrow J^{'}$, $F^{'}$ where the $J^{'}$ and $F^{'}$ quantum numbers refer to the final energy levels, and $J$ and $F$ indicate initial energy levels, collisional coefficient values with units of cm$^{3}$ s$^{-1}$ are given for five temperatures; 10, 20, 30, 40 and 50 K (Table \ref{collcoeft}).

\begin{table*}[h]
    \centering
    \caption{\hcopp \ hyperfine collisonal rate coefficients given in units of cm$^{3}$ s$^{-1}$ for 10, 20, 30, 40 and 50 K. Transitions are labelled with the $J$, $F \rightarrow J^{'}$,$F^{'}$ quantum numbers where the $J^{'}$ and $F^{'}$ quantum numbers refer to the final energy levels, and $J$ and $F$ indicate initial energy levels.}
    \label{collcoeft}
\begin{tabular}{llllll}
    \hline
    \hline
    Quantum Numbers & & &T& & \\
    \hline
    $J$, $F \rightarrow J^{'}$ ,$F^{'}$ & 10 K & 20 K & 30 K & 40 K & 50 K \\
    \hline

        0, 5/2 $\rightarrow$ 1, 3/2 & 2.390000$ \times10^{-10}$   & 2.145000$ \times10^{-10}$   & 2.037000$ \times10^{-10}$   & 1.985000$ \times10^{-10}$   & 1.962000$ \times10^{-10}$ \\  
        0, 5/2 $\rightarrow$ 1, 5/2 & 2.390000$ \times10^{-10}$   & 2.145000$ \times10^{-10}$   & 2.037000$ \times10^{-10}$   & 1.985000$ \times10^{-10}$   & 1.962000$ \times10^{-10}$ \\  
        1, 3/2 $\rightarrow$ 1, 5/2 & 3.871328$ \times10^{-11}$   & 3.444551$ \times10^{-11}$   & 3.146681$ \times10^{-11}$   & 2.914492$ \times10^{-11}$   & 2.729761$ \times10^{-11}$ \\  
        1, 7/2 $\rightarrow$ 1, 5/2 & 1.659141$ \times10^{-10}$   & 1.476236$ \times10^{-10}$   & 1.348578$ \times10^{-10}$   & 1.249068$ \times10^{-10}$   & 1.169897$ \times10^{-10}$ \\  
        0, 5/2 $\rightarrow$ 1, 7/2 & 2.390000$ \times10^{-10}$   & 2.145000$ \times10^{-10}$   & 2.037000$ \times10^{-10}$   & 1.985000$ \times10^{-10}$   & 1.962000$ \times10^{-10}$ \\  
        1, 3/2 $\rightarrow$ 1, 7/2 & 1.290443$ \times10^{-10}$   & 1.148184$ \times10^{-10}$   & 1.048894$ \times10^{-10}$   & 9.714972$ \times10^{-11}$   & 9.099203$ \times10^{-11}$ \\
\end{tabular}
\end{table*}

\begin{table*}[h]
    \centering
    \caption{Continuation of Table \ref{collcoeft}.}
    \begin{tabular}{llllll}
    \hline
    \hline
    Quantum Numbers & & &T& & \\
    \hline
    $J$, $F \rightarrow J^{'}$ ,$F^{'}$ & 10 K & 20 K & 30 K & 40 K & 50 K \\
    \hline
    
        0, 5/2  $\rightarrow$ 3, 5/2 & 9.659000$ \times10^{-11}$   & 8.195000$ \times10^{-11}$   & 7.271000$ \times10^{-11}$   & 6.659000$ \times10^{-11}$   & 6.225000$ \times10^{-11}$ \\ 
        1, 3/2  $\rightarrow$ 3, 5/2 & 9.045786$ \times10^{-11}$   & 7.823060$ \times10^{-11}$   & 7.147759$ \times10^{-11}$   & 6.679067$ \times10^{-11}$   & 6.317294$ \times10^{-11}$ \\ 
        1, 5/2  $\rightarrow$ 3, 5/2 & 1.095869$ \times10^{-10}$   & 9.597244$ \times10^{-11}$   & 8.780751$ \times10^{-11}$   & 8.207587$ \times10^{-11}$   & 7.766319$ \times10^{-11}$ \\ 
        1, 7/2  $\rightarrow$ 3, 5/2 & 7.945528$ \times10^{-11}$   & 7.239696$ \times10^{-11}$   & 6.651489$ \times10^{-11}$   & 6.223346$ \times10^{-11}$   & 5.896387$ \times10^{-11}$ \\ 
        2, 1/2  $\rightarrow$ 3, 5/2 & 1.457867$ \times10^{-11}$   & 1.286091$ \times10^{-11}$   & 1.162068$ \times10^{-11}$   & 1.079016$ \times10^{-11}$   & 1.020176$ \times10^{-11}$ \\ 
        2, 3/2  $\rightarrow$ 3, 5/2 & 1.553198$ \times10^{-10}$   & 1.434066$ \times10^{-10}$   & 1.370036$ \times10^{-10}$   & 1.339313$ \times10^{-10}$   & 1.326796$ \times10^{-10}$ \\
\end{tabular}
\end{table*}

\begin{table*}[h]
    \centering
    \caption{Continuation of Table \ref{collcoeft}.}
    \begin{tabular}{llllll}
\hline
    \hline
    Quantum Numbers & & &T& & \\
    \hline
    $J$, $F \rightarrow J^{'}$ ,$F^{'}$ & 10 K & 20 K & 30 K & 40 K & 50 K \\
    \hline
         2, 5/2 $\rightarrow$ 4, 3/2 & 6.075598$ \times10^{-11}$   & 5.606351$ \times10^{-11}$   & 5.136840$ \times10^{-11}$   & 4.790023$ \times10^{-11}$   & 4.529028$ \times10^{-11}$ \\ 
         2, 7/2 $\rightarrow$ 4, 3/2 & 2.025485$ \times10^{-11}$   & 1.887397$ \times10^{-11}$   & 1.731073$ \times10^{-11}$   & 1.614575$ \times10^{-11}$   & 1.527075$ \times10^{-11}$ \\ 
         2, 9/2 $\rightarrow$ 4, 3/2 & 5.877659$ \times10^{-11}$   & 6.223771$ \times10^{-11}$   & 6.103703$ \times10^{-11}$   & 5.894917$ \times10^{-11}$   & 5.670042$ \times10^{-11}$ \\ 
         3, 1/2 $\rightarrow$ 4, 3/2 & 1.522410$ \times10^{-10}$   & 1.454889$ \times10^{-10}$   & 1.431467$ \times10^{-10}$   & 1.429769$ \times10^{-10}$   & 1.439454$ \times10^{-10}$ \\ 
         3, 3/2 $\rightarrow$ 4, 3/2 & 1.068840$ \times10^{-10}$   & 1.005365$ \times10^{-10}$   & 9.712121$ \times10^{-11}$   & 9.547041$ \times10^{-11}$   & 9.480810$ \times10^{-11}$ \\ 
         3, 5/2 $\rightarrow$ 4, 3/2 & 5.460310$ \times10^{-11}$   & 4.977087$ \times10^{-11}$   & 4.627438$ \times10^{-11}$   & 4.391609$ \times10^{-11}$   & 4.225004$ \times10^{-11}$ \\
\end{tabular}
\end{table*}

\begin{table*}[h]
    \centering
    \caption{Continuation of Table \ref{collcoeft}.}
    \begin{tabular}{llllll}
\hline
    \hline
    Quantum Numbers & & &T& & \\
    \hline
    $J$, $F \rightarrow J^{'}$ ,$F^{'}$ & 10 K & 20 K & 30 K & 40 K & 50 K \\
    \hline
    
        3, 3/2  $\rightarrow$ 4, 9/2 & 1.645436$ \times10^{-11}$   & 1.506369$ \times10^{-11}$   & 1.397602$ \times10^{-11}$   & 1.320777$ \times10^{-11}$   & 1.264035$ \times10^{-11}$ \\ 
        3, 5/2  $\rightarrow$ 4, 9/2 & 3.343967$ \times10^{-11}$   & 3.138070$ \times10^{-11}$   & 2.958240$ \times10^{-11}$   & 2.820373$ \times10^{-11}$   & 2.711973$ \times10^{-11}$ \\ 
        3, 7/2  $\rightarrow$ 4, 9/2 & 2.118848$ \times10^{-10}$   & 2.044106$ \times10^{-10}$   & 2.014471$ \times10^{-10}$   & 2.007949$ \times10^{-10}$   & 2.014263$ \times10^{-10}$ \\ 
        3, 9/2  $\rightarrow$ 4, 9/2 & 1.069472$ \times10^{-10}$   & 1.015944$ \times10^{-10}$   & 9.789043$ \times10^{-11}$   & 9.536810$ \times10^{-11}$   & 9.361765$ \times10^{-11}$ \\ 
        3, 11/2 $\rightarrow$ 4, 9/2  & 4.785886$ \times10^{-11}$   & 4.758193$ \times10^{-11}$   & 4.654179$ \times10^{-11}$   & 4.546006$ \times10^{-11}$   & 4.446093$ \times10^{-11}$ \\
        4, 3/2  $\rightarrow$ 4, 9/2 & 1.782512$ \times10^{-11}$   & 1.913040$ \times10^{-11}$   & 1.876123$ \times10^{-11}$   & 1.801811$ \times10^{-11}$   & 1.722513$ \times10^{-11}$ \\
\end{tabular}
\end{table*}

\begin{table*}[h]
    \centering
    \caption{Continuation of Table \ref{collcoeft}.}
    \begin{tabular}{llllll}
\hline
    \hline
    Quantum Numbers & & &T& & \\
    \hline
    $J$, $F \rightarrow J^{'}$ ,$F^{'}$ & 10 K & 20 K & 30 K & 40 K & 50 K \\
    \hline
    
         3, 5/2  $\rightarrow$ 5, 5/2 & 4.862221$ \times10^{-11}$   & 4.517789$ \times10^{-11}$   & 4.176476$ \times10^{-11}$   & 3.909341$ \times10^{-11}$   & 3.704056$ \times10^{-11}$ \\ 
         3, 7/2  $\rightarrow$ 5, 5/2 & 3.407422$ \times10^{-11}$   & 3.461841$ \times10^{-11}$   & 3.332198$ \times10^{-11}$   & 3.184881$ \times10^{-11}$   & 3.049733$ \times10^{-11}$ \\ 
         3, 9/2  $\rightarrow$ 5, 5/2 & 1.652350$ \times10^{-11}$   & 1.758971$ \times10^{-11}$   & 1.731549$ \times10^{-11}$   & 1.673873$ \times10^{-11}$   & 1.611590$ \times10^{-11}$ \\ 
         3, 11/2 $\rightarrow$ 5, 5/2  & 3.614959$ \times10^{-11}$   & 3.585899$ \times10^{-11}$   & 3.554955$ \times10^{-11}$   & 3.528427$ \times10^{-11}$   & 3.501489$ \times10^{-11}$ \\
         4, 3/2  $\rightarrow$ 5, 5/2 & 1.917424$ \times10^{-10}$   & 1.839385$ \times10^{-10}$   & 1.823983$ \times10^{-10}$   & 1.831070$ \times10^{-10}$   & 1.848756$ \times10^{-10}$ \\ 
         4, 5/2  $\rightarrow$ 5, 5/2 & 8.081082$ \times10^{-11}$   & 7.644061$ \times10^{-11}$   & 7.413205$ \times10^{-11}$   & 7.284832$ \times10^{-11}$   & 7.212740$ \times10^{-11}$ \\
\end{tabular}
\end{table*}

\begin{table*}[h]
    \centering
    \caption{Continuation of Table \ref{collcoeft}.}
    \begin{tabular}{llllll}
\hline
    \hline
    Quantum Numbers & & &T& & \\
    \hline
    $J$, $F \rightarrow J^{'}$ ,$F^{'}$ & 10 K & 20 K & 30 K & 40 K & 50 K \\
    \hline
         5, 5/2  $\rightarrow$  5, 9/2 & 2.945741$ \times10^{-11}$   & 3.055133$ \times10^{-11}$   & 2.953792$ \times10^{-11}$   & 2.818409$ \times10^{-11}$   & 2.686770$ \times10^{-11}$ \\ 
         5, 7/2  $\rightarrow$  5, 9/2 & 8.093865$ \times10^{-11}$   & 7.769268$ \times10^{-11}$   & 7.340331$ \times10^{-11}$   & 6.946155$ \times10^{-11}$   & 6.606866$ \times10^{-11}$ \\ 
         5, 13/2 $\rightarrow$  5, 9/2  & 5.388127$ \times10^{-11}$   & 5.804955$ \times10^{-11}$   & 5.781065$ \times10^{-11}$   & 5.665173$ \times10^{-11}$   & 5.532132$ \times10^{-11}$ \\
         5, 15/2 $\rightarrow$  5, 9/2  & 3.013817$ \times10^{-11}$   & 3.316016$ \times10^{-11}$   & 3.388155$ \times10^{-11}$   & 3.394655$ \times10^{-11}$   & 3.374718$ \times10^{-11}$ \\
          0, 5/2 $\rightarrow$ 5, 11/2 & 3.892000$ \times10^{-11}$   & 3.839000$ \times10^{-11}$   & 3.742000$ \times10^{-11}$   & 3.652000$ \times10^{-11}$   & 3.574000$ \times10^{-11}$ \\
          1, 3/2 $\rightarrow$ 5, 11/2 & 1.712017$ \times10^{-11}$   & 1.621430$ \times10^{-11}$   & 1.519642$ \times10^{-11}$   & 1.431434$ \times10^{-11}$   & 1.355767$ \times10^{-11}$ \\
\end{tabular}
\end{table*}

\begin{table*}[h]
    \centering
    \caption{Continuation of Table \ref{collcoeft}.}
    \begin{tabular}{llllll}
    \hline
    \hline
    Quantum Numbers & & &T& & \\
    \hline
    $J$, $F \rightarrow J^{'}$ ,$F^{'}$ & 10 K & 20 K & 30 K & 40 K & 50 K \\
    \hline
         1, 3/2  $\rightarrow$ 5,  15/2& 2.704465$ \times10^{-11}$   & 2.790316$ \times10^{-11}$   & 2.731681$ \times10^{-11}$   & 2.631067$ \times10^{-11}$   & 2.518595$ \times10^{-11}$ \\ 
         1, 5/2  $\rightarrow$ 5,  15/2& 1.641996$ \times10^{-11}$   & 1.694120$ \times10^{-11}$   & 1.658521$ \times10^{-11}$   & 1.597434$ \times10^{-11}$   & 1.529147$ \times10^{-11}$ \\ 
         1, 7/2  $\rightarrow$ 5,  15/2& 9.043539$ \times10^{-11}$   & 8.255564$ \times10^{-11}$   & 7.579799$ \times10^{-11}$   & 7.061499$ \times10^{-11}$   & 6.652257$ \times10^{-11}$ \\ 
         2, 1/2  $\rightarrow$ 5,  15/2& 1.198658$ \times10^{-11}$   & 1.164543$ \times10^{-11}$   & 1.119975$ \times10^{-11}$   & 1.067887$ \times10^{-11}$   & 1.017178$ \times10^{-11}$ \\ 
         2, 3/2  $\rightarrow$ 5,  15/2& 1.663444$ \times10^{-11}$   & 1.616100$ \times10^{-11}$   & 1.554251$ \times10^{-11}$   & 1.481966$ \times10^{-11}$   & 1.411594$ \times10^{-11}$ \\ 
         2, 5/2  $\rightarrow$ 5,  15/2& 2.213228$ \times10^{-11}$   & 2.177902$ \times10^{-11}$   & 2.114258$ \times10^{-11}$   & 2.041209$ \times10^{-11}$   & 1.971093$ \times10^{-11}$ \\
\end {tabular}
\end{table*}

\section{C$^{17}$O Hyperfine Collisional Rate Coefficients}\label{c17ocoll}
We approximated the C$^{17}$O-H$_{2}$ hyperfine collisional rate coefficients from the non-hyperfine collisional rate coefficients available in LAMDA using the $Mj$-randomisation method \citep{franz:66}. For each transition $J$, $F \rightarrow J^{'}$, $F^{'}$ where the $J^{'}$ and $F^{'}$ quantum numbers refer to the final energy levels, and $J$ and $F$ indicate initial energy levels, collisional coefficient values with units of cm$^{3}$ s$^{-1}$ are given for seven temperatures; 2, 5, 10, 20, 30, 40 and 50 K (Table \ref{c17ocollta}).

\begin{table*}[h]
\footnotesize
    \centering
    \caption{$C^{17}$O  hyperfine collisonal rate coefficients given in units of cm$^{3}$ s$^{-1}$ for 2, 5, 10, 20, 30, 40 and 50 K. Transitions are labelled with the $J$, $F \rightarrow J^{'}$,$F^{'}$ quantum numbers where the $J^{'}$ and $F^{'}$ quantum numbers refer to the final energy levels, and $J$ and $F$ indicate initial energy levels.}
    \label{c17ocollta}
\begin{tabular}{llllllll}
    \hline
    \hline
 Quantum Numbers & & &T& & \\
    \hline
    $J$, $F \rightarrow J^{'}$ ,$F^{'}$ & 2 K & 5 K & 10 K & 20 K & 30 K & 40 K & 50 K \\ 
\hline
0, 5/2  $\rightarrow$ 1, 3/2  & 2.950$\times10^{-11}$ & 3.250$\times10^{-11}$ & 3.300$\times10^{-11}$ & 3.250$\times10^{-11}$ & 3.260$\times10^{-11}$ & 3.290$\times10^{-11}$ & 3.340$\times10^{-11}$ \\
0, 5/2  $\rightarrow$ 1, 7/2  & 2.950$\times10^{-11}$ & 3.250$\times10^{-11}$ & 3.300$\times10^{-11}$ & 3.250$\times10^{-11}$ & 3.260$\times10^{-11}$ & 3.290$\times10^{-11}$ & 3.340$\times10^{-11}$ \\
0, 5/2  $\rightarrow$ 1, 5/2  & 2.950$\times10^{-11}$ & 3.250$\times10^{-11}$ & 3.300$\times10^{-11}$ & 3.250$\times10^{-11}$ & 3.260$\times10^{-11}$ & 3.290$\times10^{-11}$ & 3.340$\times10^{-11}$ \\
0, 5/2  $\rightarrow$ 2, 1/2  & 2.870$\times10^{-11}$ & 2.880$\times10^{-11}$ & 3.020$\times10^{-11}$ & 3.090$\times10^{-11}$ & 3.110$\times10^{-11}$ & 3.120$\times10^{-11}$ & 3.130$\times10^{-11}$ \\
0, 5/2  $\rightarrow$ 2, 9/2  & 2.870$\times10^{-11}$ & 2.880$\times10^{-11}$ & 3.020$\times10^{-11}$ & 3.090$\times10^{-11}$ & 3.110$\times10^{-11}$ & 3.120$\times10^{-11}$ & 3.130$\times10^{-11}$ \\
0, 5/2  $\rightarrow$ 2, 3/2  & 2.870$\times10^{-11}$ & 2.880$\times10^{-11}$ & 3.020$\times10^{-11}$ & 3.090$\times10^{-11}$ & 3.110$\times10^{-11}$ & 3.120$\times10^{-11}$ & 3.130$\times10^{-11}$ \\
\end{tabular}
\end{table*}

\begin{table*}[h]
\footnotesize
    \centering
    \caption{Continuation of Table \ref{c17ocollta}.}
    \begin{tabular}{llllllll}
    \hline
    \hline
 Quantum Numbers & & &T& & \\
    \hline
    $J$, $F \rightarrow J^{'}$ ,$F^{'}$ & 2 K & 5 K & 10 K & 20 K & 30 K & 40 K & 50 K \\ 
\hline
2, 5/2  $\rightarrow$ 3, 1/2  & 2.220$\times10^{-11}$ & 1.810$\times10^{-11}$ & 1.590$\times10^{-11}$ & 1.410$\times10^{-11}$ & 1.340$\times10^{-11}$ & 1.310$\times10^{-11}$ & 1.310$\times10^{-11}$ \\
2, 7/2  $\rightarrow$ 3, 1/2  & 2.950$\times10^{-11}$ & 2.410$\times10^{-11}$ & 2.120$\times10^{-11}$ & 1.880$\times10^{-11}$ & 1.780$\times10^{-11}$ & 1.750$\times10^{-11}$ & 1.740$\times10^{-11}$ \\
2, 1/2  $\rightarrow$ 3, 3/2  & 7.390$\times10^{-12}$ & 6.030$\times10^{-12}$ & 5.300$\times10^{-12}$ & 4.700$\times10^{-12}$ & 4.460$\times10^{-12}$ & 4.370$\times10^{-12}$ & 4.350$\times10^{-12}$ \\
2, 9/2  $\rightarrow$ 3, 3/2  & 3.690$\times10^{-11}$ & 3.020$\times10^{-11}$ & 2.650$\times10^{-11}$ & 2.350$\times10^{-11}$ & 2.230$\times10^{-11}$ & 2.190$\times10^{-11}$ & 2.180$\times10^{-11}$ \\
2, 3/2  $\rightarrow$ 3, 3/2  & 1.480$\times10^{-11}$ & 1.210$\times10^{-11}$ & 1.060$\times10^{-11}$ & 9.400$\times10^{-12}$ & 8.920$\times10^{-12}$ & 8.740$\times10^{-12}$ & 8.710$\times10^{-12}$ \\
2, 5/2  $\rightarrow$ 3, 3/2  & 2.220$\times10^{-11}$ & 1.810$\times10^{-11}$ & 1.590$\times10^{-11}$ & 1.410$\times10^{-11}$ & 1.340$\times10^{-11}$ & 1.310$\times10^{-11}$ & 1.310$\times10^{-11}$ \\
\end{tabular}
\end{table*}

\begin{table*}[h]
\footnotesize
    \centering
    \caption{Continuation of Table \ref{c17ocollta}.}
    \begin{tabular}{llllllll}
    \hline
    \hline
 Quantum Numbers & & &T& & \\
    \hline
    $J$, $F \rightarrow J^{'}$ ,$F^{'}$ & 2 K & 5 K & 10 K & 20 K & 30 K & 40 K & 50 K \\ 
\hline
2, 7/2  $\rightarrow$ 4,  3/2 & 1.620$\times10^{-11}$ & 1.590$\times10^{-11}$ & 1.660$\times10^{-11}$ & 1.660$\times10^{-11}$ & 1.630$\times10^{-11}$ & 1.610$\times10^{-11}$ & 1.590$\times10^{-11}$ \\
2, 1/2  $\rightarrow$ 4, 5/2  & 4.040$\times10^{-12}$ & 3.970$\times10^{-12}$ & 4.150$\times10^{-12}$ & 4.160$\times10^{-12}$ & 4.080$\times10^{-12}$ & 4.020$\times10^{-12}$ & 3.980$\times10^{-12}$ \\
2, 9/2  $\rightarrow$ 4, 5/2  & 2.020$\times10^{-11}$ & 1.980$\times10^{-11}$ & 2.080$\times10^{-11}$ & 2.080$\times10^{-11}$ & 2.040$\times10^{-11}$ & 2.010$\times10^{-11}$ & 1.990$\times10^{-11}$ \\
2, 3/2  $\rightarrow$ 4, 5/2  & 8.080$\times10^{-12}$ & 7.940$\times10^{-12}$ & 8.310$\times10^{-12}$ & 8.320$\times10^{-12}$ & 8.160$\times10^{-12}$ & 8.040$\times10^{-12}$ & 7.960$\times10^{-12}$ \\
2, 5/2  $\rightarrow$ 4, 5/2  & 1.210$\times10^{-11}$ & 1.190$\times10^{-11}$ & 1.250$\times10^{-11}$ & 1.250$\times10^{-11}$ & 1.220$\times10^{-11}$ & 1.210$\times10^{-11}$ & 1.190$\times10^{-11}$ \\
2, 7/2  $\rightarrow$ 4, 5/2  & 1.620$\times10^{-11}$ & 1.590$\times10^{-11}$ & 1.660$\times10^{-11}$ & 1.660$\times10^{-11}$ & 1.630$\times10^{-11}$ & 1.610$\times10^{-11}$ & 1.590$\times10^{-11}$ \\
\end{tabular}
\end{table*}

\begin{table*}[h]
\footnotesize
    \centering
    \caption{Continuation of Table \ref{c17ocollta}.}
    \begin{tabular}{llllllll}
    \hline
    \hline
 Quantum Numbers & & &T& & \\
    \hline
    $J$, $F \rightarrow J^{'}$ ,$F^{'}$ & 2 K & 5 K & 10 K & 20 K & 30 K & 40 K & 50 K \\ 
    \hline
3, 9/2  $\rightarrow$ 4, 9/2  & 1.960$\times10^{-11}$ & 2.020$\times10^{-11}$ & 1.930$\times10^{-11}$ & 1.750$\times10^{-11}$ & 1.660$\times10^{-11}$ & 1.620$\times10^{-11}$ & 1.610$\times10^{-11}$ \\
3, 7/2  $\rightarrow$ 4, 9/2  & 1.570$\times10^{-11}$ & 1.620$\times10^{-11}$ & 1.550$\times10^{-11}$ & 1.400$\times10^{-11}$ & 1.330$\times10^{-11}$ & 1.300$\times10^{-11}$ & 1.290$\times10^{-11}$ \\
0, 5/2  $\rightarrow$ 5, 15/2 & 7.920$\times10^{-13}$ & 8.220$\times10^{-13}$ & 9.700$\times10^{-13}$ & 1.180$\times10^{-12}$ & 1.380$\times10^{-12}$ & 1.590$\times10^{-12}$ & 1.830$\times10^{-12}$ \\
0, 5/2  $\rightarrow$ 5, 5/2  & 7.920$\times10^{-13}$ & 8.220$\times10^{-13}$ & 9.700$\times10^{-13}$ & 1.180$\times10^{-12}$ & 1.380$\times10^{-12}$ & 1.590$\times10^{-12}$ & 1.830$\times10^{-12}$ \\
0, 5/2  $\rightarrow$ 5, 7/2  & 7.920$\times10^{-13}$ & 8.220$\times10^{-13}$ & 9.700$\times10^{-13}$ & 1.180$\times10^{-12}$ & 1.380$\times10^{-12}$ & 1.590$\times10^{-12}$ & 1.830$\times10^{-12}$ \\
0, 5/2  $\rightarrow$ 5, 13/2 & 7.920$\times10^{-13}$ & 8.220$\times10^{-13}$ & 9.700$\times10^{-13}$ & 1.180$\times10^{-12}$ & 1.380$\times10^{-12}$ & 1.590$\times10^{-12}$ & 1.830$\times10^{-12}$ \\
\end{tabular}
\end{table*}

\begin{table*}[h]
\footnotesize
    \centering
    \caption{Continuation of Table \ref{c17ocollta}.}
    \begin{tabular}{llllllll}
    \hline
    \hline
 Quantum Numbers & & &T& & \\
    \hline
    $J$, $F \rightarrow J^{'}$ ,$F^{'}$ & 2 K & 5 K & 10 K & 20 K & 30 K & 40 K & 50 K \\ 
\hline
2, 7/2  $\rightarrow$ 5, 11/2 & 2.160$\times10^{-12}$ & 2.640$\times10^{-12}$ & 3.340$\times10^{-12}$ & 3.900$\times10^{-12}$ & 4.220$\times10^{-12}$ & 4.520$\times10^{-12}$ & 4.830$\times10^{-12}$ \\
3, 11/2 $\rightarrow$ 5, 15/2 & 2.360$\times10^{-11}$ & 2.210$\times10^{-11}$ & 2.180$\times10^{-11}$ & 2.090$\times10^{-11}$ & 2.020$\times10^{-11}$ & 1.960$\times10^{-11}$ & 1.910$\times10^{-11}$ \\
3, 1/2  $\rightarrow$ 5, 15/2 & 3.940$\times10^{-12}$ & 3.680$\times10^{-12}$ & 3.630$\times10^{-12}$ & 3.490$\times10^{-12}$ & 3.360$\times10^{-12}$ & 3.260$\times10^{-12}$ & 3.190$\times10^{-12}$ \\
3, 3/2  $\rightarrow$ 5, 15/2 & 7.880$\times10^{-12}$ & 7.360$\times10^{-12}$ & 7.250$\times10^{-12}$ & 6.980$\times10^{-12}$ & 6.720$\times10^{-12}$ & 6.520$\times10^{-12}$ & 6.380$\times10^{-12}$ \\
3, 5/2  $\rightarrow$ 5, 15/2 & 1.180$\times10^{-11}$ & 1.100$\times10^{-11}$ & 1.090$\times10^{-11}$ & 1.050$\times10^{-11}$ & 1.010$\times10^{-11}$ & 9.780$\times10^{-12}$ & 9.570$\times10^{-12}$ \\
3, 9/2  $\rightarrow$ 5, 15/2 & 1.970$\times10^{-11}$ & 1.840$\times10^{-11}$ & 1.810$\times10^{-11}$ & 1.740$\times10^{-11}$ & 1.680$\times10^{-11}$ & 1.630$\times10^{-11}$ & 1.600$\times10^{-11}$ \\
\end{tabular}
\end{table*}

\begin{table*}[h]
\footnotesize
    \centering
    \caption{Continuation of Table \ref{c17ocollta}.}
    \begin{tabular}{llllllll}
    \hline
    \hline
 Quantum Numbers & & &T& & \\
    \hline
    $J$, $F \rightarrow J^{'}$ ,$F^{'}$ & 2 K & 5 K & 10 K & 20 K & 30 K & 40 K & 50 K \\ 
\hline
4, 13/2 $\rightarrow$ 5, 13/2 & 6.120$\times10^{-12}$ & 1.260$\times10^{-11}$ & 1.700$\times10^{-11}$ & 1.820$\times10^{-11}$ & 1.810$\times10^{-11}$ & 1.800$\times10^{-11}$ & 1.800$\times10^{-11}$ \\
4,  3/2 $\rightarrow$ 5, 13/2 & 1.750$\times10^{-12}$ & 3.590$\times10^{-12}$ & 4.850$\times10^{-12}$ & 5.210$\times10^{-12}$ & 5.170$\times10^{-12}$ & 5.140$\times10^{-12}$ & 5.130$\times10^{-12}$ \\
4, 5/2  $\rightarrow$ 5, 13/2 & 2.620$\times10^{-12}$ & 5.390$\times10^{-12}$ & 7.280$\times10^{-12}$ & 7.820$\times10^{-12}$ & 7.760$\times10^{-12}$ & 7.700$\times10^{-12}$ & 7.700$\times10^{-12}$ \\
4, 11/2 $\rightarrow$ 5, 13/2 & 5.240$\times10^{-12}$ & 1.080$\times10^{-11}$ & 1.460$\times10^{-11}$ & 1.560$\times10^{-11}$ & 1.550$\times10^{-11}$ & 1.540$\times10^{-11}$ & 1.540$\times10^{-11}$ \\
4, 7/2  $\rightarrow$ 5, 13/2 & 3.490$\times10^{-12}$ & 7.180$\times10^{-12}$ & 9.710$\times10^{-12}$ & 1.040$\times10^{-11}$ & 1.030$\times10^{-11}$ & 1.030$\times10^{-11}$ & 1.030$\times10^{-11}$ \\
4, 9/2  $\rightarrow$ 5, 13/2 & 4.370$\times10^{-12}$ & 8.980$\times10^{-12}$ & 1.210$\times10^{-11}$ & 1.300$\times10^{-11}$ & 1.290$\times10^{-11}$ & 1.280$\times10^{-11}$ & 1.280$\times10^{-11}$ \\
\end{tabular}
\end{table*}

\begin{table*}[h]
\footnotesize
    \centering
    \caption{Continuation of Table \ref{c17ocollta}.}
    \begin{tabular}{llllllll}
    \hline
    \hline
 Quantum Numbers & & &T& & \\
    \hline
    $J$, $F \rightarrow J^{'}$ ,$F^{'}$ & 2 K & 5 K & 10 K & 20 K & 30 K & 40 K & 50 K \\ 
\hline
2, 5/2  $\rightarrow$ 6, 15/2 & 1.980$\times10^{-12}$ & 1.830$\times10^{-12}$ & 1.780$\times10^{-12}$ & 1.780$\times10^{-12}$ & 1.810$\times10^{-12}$ & 1.850$\times10^{-12}$ & 1.900$\times10^{-12}$ \\
2, 7/2  $\rightarrow$ 6, 15/2 & 2.640$\times10^{-12}$ & 2.440$\times10^{-12}$ & 2.380$\times10^{-12}$ & 2.380$\times10^{-12}$ & 2.410$\times10^{-12}$ & 2.470$\times10^{-12}$ & 2.530$\times10^{-12}$ \\
2, 1/2  $\rightarrow$ 6, 9/2  & 6.590$\times10^{-13}$ & 6.100$\times10^{-13}$ & 5.950$\times10^{-13}$ & 5.950$\times10^{-13}$ & 6.040$\times10^{-13}$ & 6.170$\times10^{-13}$ & 6.320$\times10^{-13}$ \\
2, 9/2  $\rightarrow$ 6, 9/2  & 3.290$\times10^{-12}$ & 3.050$\times10^{-12}$ & 2.970$\times10^{-12}$ & 2.970$\times10^{-12}$ & 3.020$\times10^{-12}$ & 3.080$\times10^{-12}$ & 3.160$\times10^{-12}$ \\
2, 3/2  $\rightarrow$ 6, 9/2  & 1.320$\times10^{-12}$ & 1.220$\times10^{-12}$ & 1.190$\times10^{-12}$ & 1.190$\times10^{-12}$ & 1.210$\times10^{-12}$ & 1.230$\times10^{-12}$ & 1.260$\times10^{-12}$ \\
2, 5/2  $\rightarrow$ 6, 9/2  & 1.980$\times10^{-12}$ & 1.830$\times10^{-12}$ & 1.780$\times10^{-12}$ & 1.780$\times10^{-12}$ & 1.810$\times10^{-12}$ & 1.850$\times10^{-12}$ & 1.900$\times10^{-12}$ \\
\end{tabular}
\end{table*}

\begin{table*}[h]
\footnotesize
    \centering
    \caption{Continuation of Table \ref{c17ocollta}.}
    \begin{tabular}{llllllll}
    \hline
    \hline
 Quantum Numbers & & &T& & \\
    \hline
    $J$, $F \rightarrow J^{'}$ ,$F^{'}$ & 2 K & 5 K & 10 K & 20 K & 30 K & 40 K & 50 K \\ 
    \hline
4,  3/2 $\rightarrow$ 6, 17/2 & 6.470$\times10^{-12}$ & 6.370$\times10^{-12}$ & 6.330$\times10^{-12}$ & 6.040$\times10^{-12}$ & 5.760$\times10^{-12}$ & 5.550$\times10^{-12}$ & 5.390$\times10^{-12}$ \\
4, 5/2  $\rightarrow$ 6, 17/2 & 9.710$\times10^{-12}$ & 9.550$\times10^{-12}$ & 9.500$\times10^{-12}$ & 9.060$\times10^{-12}$ & 8.640$\times10^{-12}$ & 8.320$\times10^{-12}$ & 8.090$\times10^{-12}$ \\
4, 11/2 $\rightarrow$ 6, 17/2 & 1.940$\times10^{-11}$ & 1.910$\times10^{-11}$ & 1.900$\times10^{-11}$ & 1.810$\times10^{-11}$ & 1.730$\times10^{-11}$ & 1.660$\times10^{-11}$ & 1.620$\times10^{-11}$ \\
4, 7/2  $\rightarrow$ 6, 17/2 & 1.290$\times10^{-11}$ & 1.270$\times10^{-11}$ & 1.270$\times10^{-11}$ & 1.210$\times10^{-11}$ & 1.150$\times10^{-11}$ & 1.110$\times10^{-11}$ & 1.080$\times10^{-11}$ \\
4, 9/2  $\rightarrow$ 6, 17/2 & 1.620$\times10^{-11}$ & 1.590$\times10^{-11}$ & 1.580$\times10^{-11}$ & 1.510$\times10^{-11}$ & 1.440$\times10^{-11}$ & 1.390$\times10^{-11}$ & 1.350$\times10^{-11}$ \\
4, 13/2 $\rightarrow$ 6, 7/2  & 2.270$\times10^{-11}$ & 2.230$\times10^{-11}$ & 2.220$\times10^{-11}$ & 2.110$\times10^{-11}$ & 2.020$\times10^{-11}$ & 1.940$\times10^{-11}$ & 1.890$\times10^{-11}$ \\
\end{tabular}
\end{table*}

\begin{table*}[h]
\footnotesize
    \centering
    \caption{Continuation of Table \ref{c17ocollta}.}
    \begin{tabular}{llllllll}
    \hline
    \hline
 Quantum Numbers & & &T& & \\
    \hline
    $J$, $F \rightarrow J^{'}$ ,$F^{'}$ & 2 K & 5 K & 10 K & 20 K & 30 K & 40 K & 50 K \\ 
\hline
5, 13/2 $\rightarrow$ 6, 9/2  & 5.760$\times10^{-12}$ & 8.420$\times10^{-12}$ & 1.230$\times10^{-11}$ & 1.460$\times10^{-11}$ & 1.500$\times10^{-11}$ & 1.510$\times10^{-11}$ & 1.510$\times10^{-11}$ \\
5, 9/2  $\rightarrow$ 6, 9/2  & 4.120$\times10^{-12}$ & 6.010$\times10^{-12}$ & 8.790$\times10^{-12}$ & 1.040$\times10^{-11}$ & 1.070$\times10^{-11}$ & 1.080$\times10^{-11}$ & 1.080$\times10^{-11}$ \\
5, 11/2 $\rightarrow$ 6, 9/2  & 4.940$\times10^{-12}$ & 7.220$\times10^{-12}$ & 1.050$\times10^{-11}$ & 1.250$\times10^{-11}$ & 1.280$\times10^{-11}$ & 1.290$\times10^{-11}$ & 1.300$\times10^{-11}$ \\
5, 15/2 $\rightarrow$ 6, 11/2 & 6.590$\times10^{-12}$ & 9.620$\times10^{-12}$ & 1.410$\times10^{-11}$ & 1.670$\times10^{-11}$ & 1.710$\times10^{-11}$ & 1.720$\times10^{-11}$ & 1.730$\times10^{-11}$ \\
5, 5/2  $\rightarrow$ 6, 11/2 & 2.470$\times10^{-12}$ & 3.610$\times10^{-12}$ & 5.270$\times10^{-12}$ & 6.260$\times10^{-12}$ & 6.420$\times10^{-12}$ & 6.450$\times10^{-12}$ & 6.480$\times10^{-12}$ \\
5, 7/2  $\rightarrow$ 6, 11/2 & 3.290$\times10^{-12}$ & 4.810$\times10^{-12}$ & 7.030$\times10^{-12}$ & 8.350$\times10^{-12}$ & 8.560$\times10^{-12}$ & 8.600$\times10^{-12}$ & 8.640$\times10^{-12}$ \\
\end{tabular}
\end{table*}

\begin{table*}[h]
\footnotesize
    \centering
    \caption{Continuation of Table \ref{c17ocollta}.}
    \begin{tabular}{llllllll}
    \hline
    \hline
 Quantum Numbers & & &T& & \\
    \hline
    $J$, $F \rightarrow J^{'}$ ,$F^{'}$ & 2 K & 5 K & 10 K & 20 K & 30 K & 40 K & 50 K \\ 
    \hline
2, 9/2  $\rightarrow$ 7, 11/2 & 1.210$\times10^{-12}$ & 1.160$\times10^{-12}$ & 1.200$\times10^{-12}$ & 1.320$\times10^{-12}$ & 1.450$\times10^{-12}$ & 1.600$\times10^{-12}$ & 1.770$\times10^{-12}$ \\
2, 3/2  $\rightarrow$ 7, 11/2 & 4.830$\times10^{-13}$ & 4.650$\times10^{-13}$ & 4.810$\times10^{-13}$ & 5.280$\times10^{-13}$ & 5.810$\times10^{-13}$ & 6.410$\times10^{-13}$ & 7.060$\times10^{-13}$ \\
2, 5/2  $\rightarrow$ 7, 11/2 & 7.250$\times10^{-13}$ & 6.980$\times10^{-13}$ & 7.210$\times10^{-13}$ & 7.930$\times10^{-13}$ & 8.720$\times10^{-13}$ & 9.610$\times10^{-13}$ & 1.060$\times10^{-12}$ \\
2, 7/2  $\rightarrow$ 7, 11/2 & 9.660$\times10^{-13}$ & 9.310$\times10^{-13}$ & 9.610$\times10^{-13}$ & 1.060$\times10^{-12}$ & 1.160$\times10^{-12}$ & 1.280$\times10^{-12}$ & 1.410$\times10^{-12}$ \\
2, 1/2  $\rightarrow$ 7, 15/2 & 2.420$\times10^{-13}$ & 2.330$\times10^{-13}$ & 2.400$\times10^{-13}$ & 2.640$\times10^{-13}$ & 2.910$\times10^{-13}$ & 3.200$\times10^{-13}$ & 3.530$\times10^{-13}$ \\
2, 9/2  $\rightarrow$ 7, 15/2 & 1.210$\times10^{-12}$ & 1.160$\times10^{-12}$ & 1.200$\times10^{-12}$ & 1.320$\times10^{-12}$ & 1.450$\times10^{-12}$ & 1.600$\times10^{-12}$ & 1.770$\times10^{-12}$ \\
\end{tabular}
\end{table*}

\begin{table*}[h]
\footnotesize
    \centering
    \caption{Continuation of Table \ref{c17ocollta}.}
    \begin{tabular}{llllllll}
    \hline
    \hline
 Quantum Numbers & & &T& & \\
    \hline
    $J$, $F \rightarrow J^{'}$ ,$F^{'}$ & 2 K & 5 K & 10 K & 20 K & 30 K & 40 K & 50 K \\ 
\hline
4, 9/2  $\rightarrow$ 7, 19/2 & 3.870$\times10^{-12}$ & 3.810$\times10^{-12}$ & 4.050$\times10^{-12}$ & 4.410$\times10^{-12}$ & 4.640$\times10^{-12}$ & 4.830$\times10^{-12}$ & 5.020$\times10^{-12}$ \\
4, 13/2 $\rightarrow$ 7, 9/2  & 5.420$\times10^{-12}$ & 5.330$\times10^{-12}$ & 5.670$\times10^{-12}$ & 6.180$\times10^{-12}$ & 6.490$\times10^{-12}$ & 6.760$\times10^{-12}$ & 7.030$\times10^{-12}$ \\
4,  3/2 $\rightarrow$ 7, 9/2  & 1.550$\times10^{-12}$ & 1.520$\times10^{-12}$ & 1.620$\times10^{-12}$ & 1.770$\times10^{-12}$ & 1.850$\times10^{-12}$ & 1.930$\times10^{-12}$ & 2.010$\times10^{-12}$ \\
4, 5/2  $\rightarrow$ 7, 9/2  & 2.320$\times10^{-12}$ & 2.290$\times10^{-12}$ & 2.430$\times10^{-12}$ & 2.650$\times10^{-12}$ & 2.780$\times10^{-12}$ & 2.900$\times10^{-12}$ & 3.010$\times10^{-12}$ \\
4, 11/2 $\rightarrow$ 7, 9/2  & 4.650$\times10^{-12}$ & 4.570$\times10^{-12}$ & 4.860$\times10^{-12}$ & 5.300$\times10^{-12}$ & 5.560$\times10^{-12}$ & 5.800$\times10^{-12}$ & 6.020$\times10^{-12}$ \\
4, 7/2  $\rightarrow$ 7, 9/2  & 3.100$\times10^{-12}$ & 3.050$\times10^{-12}$ & 3.240$\times10^{-12}$ & 3.530$\times10^{-12}$ & 3.710$\times10^{-12}$ & 3.870$\times10^{-12}$ & 4.020$\times10^{-12}$ \\
\end{tabular}
\end{table*}

\begin{table*}[h]
\footnotesize
    \centering
    \caption{Continuation of Table \ref{c17ocollta}.}
    \begin{tabular}{llllllll}
    \hline
    \hline
 Quantum Numbers & & &T& & \\
    \hline
    $J$, $F \rightarrow J^{'}$ ,$F^{'}$ & 2 K & 5 K & 10 K & 20 K & 30 K & 40 K & 50 K \\ 
    \hline
5, 15/2 $\rightarrow$ 7, 15/2 & 2.090$\times10^{-11}$ & 2.130$\times10^{-11}$ & 2.180$\times10^{-11}$ & 2.100$\times10^{-11}$ & 2.000$\times10^{-11}$ & 1.930$\times10^{-11}$ & 1.870$\times10^{-11}$ \\
5, 5/2  $\rightarrow$ 7, 15/2 & 7.830$\times10^{-12}$ & 7.980$\times10^{-12}$ & 8.170$\times10^{-12}$ & 7.870$\times10^{-12}$ & 7.520$\times10^{-12}$ & 7.230$\times10^{-12}$ & 7.010$\times10^{-12}$ \\
5, 7/2  $\rightarrow$ 7, 15/2 & 1.040$\times10^{-11}$ & 1.060$\times10^{-11}$ & 1.090$\times10^{-11}$ & 1.050$\times10^{-11}$ & 1.000$\times10^{-11}$ & 9.640$\times10^{-12}$ & 9.350$\times10^{-12}$ \\
5, 13/2 $\rightarrow$ 7, 15/2 & 1.830$\times10^{-11}$ & 1.860$\times10^{-11}$ & 1.910$\times10^{-11}$ & 1.840$\times10^{-11}$ & 1.750$\times10^{-11}$ & 1.690$\times10^{-11}$ & 1.640$\times10^{-11}$ \\
5, 9/2  $\rightarrow$ 7, 15/2 & 1.310$\times10^{-11}$ & 1.330$\times10^{-11}$ & 1.360$\times10^{-11}$ & 1.310$\times10^{-11}$ & 1.250$\times10^{-11}$ & 1.200$\times10^{-11}$ & 1.170$\times10^{-11}$ \\
5, 11/2 $\rightarrow$ 7, 15/2 & 1.570$\times10^{-11}$ & 1.600$\times10^{-11}$ & 1.630$\times10^{-11}$ & 1.570$\times10^{-11}$ & 1.500$\times10^{-11}$ & 1.450$\times10^{-11}$ & 1.400$\times10^{-11}$ \\
\end{tabular}
\end{table*}

\begin{table*}[h]
\footnotesize
    \centering
    \caption{Continuation of Table \ref{c17ocollta}.}
    \begin{tabular}{llllllll}
    \hline
    \hline
 Quantum Numbers & & &T& & \\
    \hline
    $J$, $F \rightarrow J^{'}$ ,$F^{'}$ & 2 K & 5 K & 10 K & 20 K & 30 K & 40 K & 50 K \\ 
\hline
1, 5/2  $\rightarrow$ 8, 21/2 & 1.810$\times10^{-13}$ & 1.770$\times10^{-13}$ & 1.830$\times10^{-13}$ & 2.020$\times10^{-13}$ & 2.280$\times10^{-13}$ & 2.630$\times10^{-13}$ & 3.030$\times10^{-13}$ \\
1, 3/2  $\rightarrow$ 8, 19/2 & 1.200$\times10^{-13}$ & 1.180$\times10^{-13}$ & 1.220$\times10^{-13}$ & 1.340$\times10^{-13}$ & 1.520$\times10^{-13}$ & 1.750$\times10^{-13}$ & 2.020$\times10^{-13}$ \\
1, 7/2  $\rightarrow$ 8, 19/2 & 2.410$\times10^{-13}$ & 2.370$\times10^{-13}$ & 2.440$\times10^{-13}$ & 2.690$\times10^{-13}$ & 3.050$\times10^{-13}$ & 3.500$\times10^{-13}$ & 4.040$\times10^{-13}$ \\
1, 5/2  $\rightarrow$ 8, 19/2 & 1.810$\times10^{-13}$ & 1.770$\times10^{-13}$ & 1.830$\times10^{-13}$ & 2.020$\times10^{-13}$ & 2.280$\times10^{-13}$ & 2.630$\times10^{-13}$ & 3.030$\times10^{-13}$ \\
1, 3/2  $\rightarrow$ 8, 11/2 & 1.200$\times10^{-13}$ & 1.180$\times10^{-13}$ & 1.220$\times10^{-13}$ & 1.340$\times10^{-13}$ & 1.520$\times10^{-13}$ & 1.750$\times10^{-13}$ & 2.020$\times10^{-13}$ \\
1, 7/2  $\rightarrow$ 8, 11/2 & 2.410$\times10^{-13}$ & 2.370$\times10^{-13}$ & 2.440$\times10^{-13}$ & 2.690$\times10^{-13}$ & 3.050$\times10^{-13}$ & 3.500$\times10^{-13}$ & 4.040$\times10^{-13}$ \\
\end{tabular}
\end{table*}

\begin{table*}[h]
\footnotesize
    \centering
    \caption{Continuation of Table \ref{c17ocollta}.}
    \begin{tabular}{llllllll}
    \hline
    \hline
 Quantum Numbers & & &T& & \\
    \hline
    $J$, $F \rightarrow J^{'}$ ,$F^{'}$ & 2 K & 5 K & 10 K & 20 K & 30 K & 40 K & 50 K \\ 
\hline
3, 7/2  $\rightarrow$ 8, 11/2 & 1.020$\times10^{-12}$ & 9.890$\times10^{-13}$ & 1.000$\times10^{-12}$ & 1.060$\times10^{-12}$ & 1.140$\times10^{-12}$ & 1.220$\times10^{-12}$ & 1.320$\times10^{-12}$ \\
3, 11/2 $\rightarrow$ 8, 13/2 & 1.520$\times10^{-12}$ & 1.480$\times10^{-12}$ & 1.510$\times10^{-12}$ & 1.590$\times10^{-12}$ & 1.700$\times10^{-12}$ & 1.840$\times10^{-12}$ & 1.980$\times10^{-12}$ \\
3, 1/2  $\rightarrow$ 8, 13/2 & 2.540$\times10^{-13}$ & 2.470$\times10^{-13}$ & 2.510$\times10^{-13}$ & 2.650$\times10^{-13}$ & 2.840$\times10^{-13}$ & 3.060$\times10^{-13}$ & 3.300$\times10^{-13}$ \\
3, 3/2  $\rightarrow$ 8, 13/2 & 5.080$\times10^{-13}$ & 4.940$\times10^{-13}$ & 5.020$\times10^{-13}$ & 5.300$\times10^{-13}$ & 5.680$\times10^{-13}$ & 6.120$\times10^{-13}$ & 6.610$\times10^{-13}$ \\
3, 5/2  $\rightarrow$ 8, 13/2 & 7.620$\times10^{-13}$ & 7.420$\times10^{-13}$ & 7.530$\times10^{-13}$ & 7.950$\times10^{-13}$ & 8.510$\times10^{-13}$ & 9.180$\times10^{-13}$ & 9.910$\times10^{-13}$ \\
3, 9/2  $\rightarrow$ 8, 13/2 & 1.270$\times10^{-12}$ & 1.240$\times10^{-12}$ & 1.250$\times10^{-12}$ & 1.320$\times10^{-12}$ & 1.420$\times10^{-12}$ & 1.530$\times10^{-12}$ & 1.650$\times10^{-12}$ \\
\end{tabular}
\end{table*}

\begin{table*}[h]
\footnotesize
    \centering
    \caption{Continuation of Table \ref{c17ocollta}.}
    \begin{tabular}{llllllll}
    \hline
    \hline
 Quantum Numbers & & &T& & \\
    \hline
    $J$, $F \rightarrow J^{'}$ ,$F^{'}$ & 2 K & 5 K & 10 K & 20 K & 30 K & 40 K & 50 K \\ 
\hline
5, 15/2 $\rightarrow$ 8, 11/2 & 6.540$\times10^{-12}$ & 6.500$\times10^{-12}$ & 6.700$\times10^{-12}$ & 7.020$\times10^{-12}$ & 7.240$\times10^{-12}$ & 7.450$\times10^{-12}$ & 7.660$\times10^{-12}$ \\
5, 5/2  $\rightarrow$ 8, 11/2 & 2.450$\times10^{-12}$ & 2.440$\times10^{-12}$ & 2.510$\times10^{-12}$ & 2.630$\times10^{-12}$ & 2.720$\times10^{-12}$ & 2.790$\times10^{-12}$ & 2.870$\times10^{-12}$ \\
5, 7/2  $\rightarrow$ 8, 11/2 & 3.270$\times10^{-12}$ & 3.250$\times10^{-12}$ & 3.350$\times10^{-12}$ & 3.510$\times10^{-12}$ & 3.620$\times10^{-12}$ & 3.730$\times10^{-12}$ & 3.830$\times10^{-12}$ \\
5, 13/2 $\rightarrow$ 8, 11/2 & 5.720$\times10^{-12}$ & 5.690$\times10^{-12}$ & 5.860$\times10^{-12}$ & 6.140$\times10^{-12}$ & 6.340$\times10^{-12}$ & 6.520$\times10^{-12}$ & 6.710$\times10^{-12}$ \\
5, 9/2  $\rightarrow$ 8, 11/2 & 4.090$\times10^{-12}$ & 4.060$\times10^{-12}$ & 4.190$\times10^{-12}$ & 4.390$\times10^{-12}$ & 4.530$\times10^{-12}$ & 4.660$\times10^{-12}$ & 4.790$\times10^{-12}$ \\
5, 11/2 $\rightarrow$ 8, 11/2 & 4.910$\times10^{-12}$ & 4.870$\times10^{-12}$ & 5.030$\times10^{-12}$ & 5.260$\times10^{-12}$ & 5.430$\times10^{-12}$ & 5.590$\times10^{-12}$ & 5.750$\times10^{-12}$ \\
\end{tabular}
\end{table*}

\begin{table*}[h]
\footnotesize
    \centering
    \caption{Continuation of Table \ref{c17ocollta}.}
    \begin{tabular}{llllllll}
    \hline
    \hline
 Quantum Numbers & & &T& & \\
    \hline
    $J$, $F \rightarrow J^{'}$ ,$F^{'}$ & 2 K & 5 K & 10 K & 20 K & 30 K & 40 K & 50 K \\
    \hline
7, 9/2  $\rightarrow$ 8, 19/2 & 5.370$\times10^{-12}$ & 5.950$\times10^{-12}$ & 7.020$\times10^{-12}$ & 8.130$\times10^{-12}$ & 8.480$\times10^{-12}$ & 8.600$\times10^{-12}$ & 8.650$\times10^{-12}$ \\
7, 17/2 $\rightarrow$ 8, 19/2 & 9.670$\times10^{-12}$ & 1.070$\times10^{-11}$ & 1.260$\times10^{-11}$ & 1.460$\times10^{-11}$ & 1.530$\times10^{-11}$ & 1.550$\times10^{-11}$ & 1.560$\times10^{-11}$ \\
7, 11/2 $\rightarrow$ 8, 19/2 & 6.450$\times10^{-12}$ & 7.140$\times10^{-12}$ & 8.420$\times10^{-12}$ & 9.760$\times10^{-12}$ & 1.020$\times10^{-11}$ & 1.030$\times10^{-11}$ & 1.040$\times10^{-11}$ \\
7, 15/2 $\rightarrow$ 8, 19/2 & 8.600$\times10^{-12}$ & 9.520$\times10^{-12}$ & 1.120$\times10^{-11}$ & 1.300$\times10^{-11}$ & 1.360$\times10^{-11}$ & 1.380$\times10^{-11}$ & 1.380$\times10^{-11}$ \\
7, 13/2 $\rightarrow$ 8, 19/2 & 7.520$\times10^{-12}$ & 8.330$\times10^{-12}$ & 9.820$\times10^{-12}$ & 1.140$\times10^{-11}$ & 1.190$\times10^{-11}$ & 1.200$\times10^{-11}$ & 1.210$\times10^{-11}$ \\
7, 19/2 $\rightarrow$ 8, 11/2 & 1.070$\times10^{-11}$ & 1.190$\times10^{-11}$ & 1.400$\times10^{-11}$ & 1.630$\times10^{-11}$ & 1.700$\times10^{-11}$ & 1.720$\times10^{-11}$ & 1.730$\times10^{-11}$ \\
\end{tabular}
\end{table*}

\end{document}